\pdfoutput=1
\documentclass[cernpreprint,texlive=2016,txfonts,UKenglish]{latex/atlasdoc}

\usepackage{latex/atlaspackage}
\usepackage{latex/atlasphysics}
\usepackage{multirow}
\usepackage{amsmath}
\usepackage{hyperref}
\usepackage{float}
\usepackage{pdflscape}
\usepackage{longtable}
\usepackage{latex/atlasbiblatex}

\usepackage{latex/atlascontribute}

\graphicspath{{logos/}}

\usepackage{lldefs}
\usepackage{bm}

\AtlasTitle{Search for new high-mass phenomena in the dilepton final state using 36~\ifb\ of proton--proton collision data at $\sqrt{s}~=$~13~\TeV{} with the ATLAS detector}

\author{The ATLAS Collaboration}

\PreprintIdNumber{CERN-EP-2017-119}

\AtlasJournal{JHEP}
\AtlasDOI{10.1007/JHEP10(2017)182}

\AtlasAbstract{%
A search is conducted for new resonant and non-resonant high-mass phenomena in dielectron and dimuon final states. The search uses \lumill\ of proton--proton collision data, collected at $\sqrt{s}$ = 13~\TeV{} by the ATLAS experiment at the LHC in 2015 and 2016. No significant deviation from the Standard Model prediction is observed. Upper limits at 95\% credibility level are set on the cross-section times branching ratio for resonances decaying into dileptons, which are converted to lower limits on the resonance mass, up to \zpllchiobs~\TeV{} for the \esix-motivated \zpchi. Lower limits on the $qq \ell\ell$ contact interaction scale are set between \CIllobsLow~\TeV{} and \CIllobsHigh~\TeV{}, depending on the model.}

\hypersetup{pdftitle={ATLAS document},pdfauthor={The ATLAS Collaboration}}

\def\thetamin{\ensuremath{\theta_{\textrm{Min}}}}

\begin{document}

\maketitle



\section{Introduction}
\label{sec:intro}
This article presents a search for resonant and non-resonant new phenomena, based on the analysis of dilepton final states ($ee$ and $\mu\mu$) in proton--proton ($pp$) collisions with the ATLAS detector at the Large Hadron Collider (LHC) operating at $\rts = 13$ \TeV{}. The data set was collected during 2015 and 2016, and corresponds to an integrated luminosity of \lumill. In the search for new physics carried out at hadron colliders, the study of dilepton final states provides excellent sensitivity to a large variety of phenomena. This experimental signature benefits from a fully reconstructed final state, high signal-selection efficiencies and relatively small, well-understood backgrounds, representing a powerful test for a wide range of theories beyond the Standard Model (SM).

Models with extended gauge groups often feature additional \u1\ symmetries with corresponding heavy spin-1 bosons. These bosons, generally referred to as \zp, would manifest as a narrow resonance through its decay, in the dilepton mass spectrum. Among these models are those inspired by Grand Unified Theories, which are motivated by gauge unification or a restoration of the left--right symmetry violated by the weak interaction. Examples considered in this article include the \zp\ bosons of the \esix-motivated~\cite{London:1986dk,Langacker:2008yv} theories as well as Minimal models~\cite{Salvioni:2009mt}.
The Sequential Standard Model (SSM)~\cite{Langacker:2008yv} is also considered due to its inherent simplicity and usefulness as a benchmark model. The SSM manifests a \zpssm\ boson with couplings to fermions equal to those of the SM $Z$ boson.

The most sensitive previous searches for a \zp\ boson decaying into the dilepton final state were carried out by the ATLAS and CMS collaborations~\cite{Aaboud:2016cth, Khachatryan:2016zqb}. Using 3.2~\ifb{} of $pp$ collision data at $\sqrt{s}$ = 13~\TeV{} collected in 2015, ATLAS set a lower exclusion limit at 95\% credibility level (CL) on the \zpssm\ pole mass of 3.4~\TeV{} for the combined $ee$ and $\mu\mu$ channels. Similar limits were set by CMS using the 2015 data sample.

This search is also sensitive to a series of other models that predict the presence of narrow dilepton resonances. These models include the Randall--Sundrum (RS) model~\cite{Randall:1999ee} with a warped extra dimension giving rise to spin-2 graviton excitations, the quantum black-hole model~\cite{Meade:2007sz}, the \zstar\ model~\cite{Chizhov:2008tp}, and the minimal walking technicolour model~\cite{Sannino:2004qp}. In order to facilitate interpretation of the results in the context of these or any other model predicting a new dilepton resonance, limits are set on the production of a generic \zp-like excess.

In addition to the search for narrow resonances, results for non-resonant phenomena are also reported. Such models of these phenomena include an effective four-fermion contact interaction (CI) between two initial-state quarks and two final-state leptons ($qq\ell\ell$). Unlike resonance models, which require sufficient energy to produce the new gauge boson, the presence of a new interaction in the non-resonant regime can be detected at a much lower energy.

The most stringent constraints from CI searches are also provided by the ATLAS and CMS collaborations~\cite{Aaboud:2016cth, Khachatryan:2014fba}, for couplings between quarks and leptons. Using 3.2~\ifb{} of $pp$ collision data at $\sqrt{s}$ = 13~\TeV{} collected in 2015, ATLAS set lower limits on the $q q \ell\ell$ CI scale of $\Lambda = 25~\TeV{}$ and $\Lambda = 18~\TeV{}$ at 95\% CL for constructive and destructive interference, respectively, in the case of left--left interactions and assuming a uniform positive prior probability in 1/$\Lambda^2$. Similar limits were set by CMS using the 2015 data set. Both the resonant and non-resonant models considered as the benchmark for this search
are further discussed in Section~\ref{sec:models}.

The presented search utilises the invariant mass spectra of the observed dilepton final states as discriminating variables. The analysis and interpretation of these spectra rely primarily on simulated samples of signal and background processes. The interpretation is performed taking into account the expected shape of different signals in the dilepton mass distribution. The use of the shape of the full dilepton invariant mass distribution reduces the uncertainties in the background modelling, thereby increasing the sensitivity of this search at high masses. This article is structured as follows: Section~\ref{sec:models} covers the theoretical motivation of the models considered in this search, followed by a description of the ATLAS detector in Section~\ref{sec:detector}, and a summary in Section~\ref{sec:data-mc} of the data and Monte Carlo (MC) samples used. The event selection is motivated and described in Section~\ref{sec:selection}, with details of the background estimation given in Section~\ref{sec:background}, and an overview of the systematic uncertainty treatment given in Section~\ref{sec:systematics}. The event yields and main kinematic distributions are presented in Section~\ref{sec:yields}, followed by a description of the statistical analysis in Section~\ref{sec:statistics}, and the results in Section~\ref{sec:result}.

\section{Theoretical models}
\label{sec:models}
\subsection{\esix -motivated \zp\ models}
\label{theory:e6}

In the class of models based on the \esix\ gauge group~\cite{London:1986dk,Langacker:2008yv},
the unified symmetry group can break to the SM in a number of different
ways.
In many of them, \esix\ is first broken to $\mathrm{SO}(10) \times \mathrm{U}(1)_\psi$,
with $\mathrm{SO}(10)$ then breaking either to
$\mathrm{SU}(4) \times \mathrm{SU}(2)_\mathrm{L} \times \mathrm{SU}(2)_\mathrm{R}$ or $\mathrm{SU}(5) \times \mathrm{U}(1)_\chi$.
In the first of these two possibilities, a \zpthreeR\ coming from $\mathrm{SU}(2)_\mathrm{R}$, where $\mathrm{3R}$ stands for the right-handed third component of weak isospin, or a \zpBL\ from the breaking of $\mathrm{SU}(4)$ into $\mathrm{SU}(3)_\mathrm{C} \times \mathrm{U}(1)_\mathrm{B-L}$
could exist at the \TeV{} scale, where B~(L) is the baryon (lepton)
number and ($\mathrm{B-L}$) is the conserved quantum number.
Both of these \zp\ bosons appear in the \MM\ discussed in the next section.
In the $\mathrm{SU}(5)$ case, the presence of $\mathrm{U}(1)_\psi$ and $\mathrm{U}(1)_\chi$ symmetries
implies the existence of associated gauge bosons \zppsi\ and \zpchi\ that can mix.
When $\mathrm{SU}(5)$ is broken down to the SM, one of the $\mathrm{U}(1)$ can remain unbroken down
to intermediate energy scales.
Therefore, the precise model is governed by a mixing angle $\te6$, with the new
potentially observable \zp\ boson defined by $\zp(\te6)=\zppsi\cos\te6+\zpchi\sin\te6$.
The value of $\te6$ specifies the \zp\ boson's coupling strength to SM fermions
as well as its intrinsic width.
In comparison to the benchmark \zpssm, which has a width of approximately 3\% of its mass,
the \esix\ models predict narrower \zp\ signals.
The \zppsi\ considered here has a width of 0.5\% of its mass, and the \zpchi\ has a
width of 1.2\% of its mass~\cite{Dittmar:2003ir,Accomando:2010fz}.
All other \zp\ signals in this model, including \zpsq, \zpI, \zpeta, and \zpN, are defined by specific values of $\te6$ ranging
from 0 to $\pi$, and have widths between those of the \zppsi\ and \zpchi.

\subsection{Minimal \zp\ models}
\label{theory:mm}

In the \MM~\cite{Salvioni:2009mt}, the phenomenology of \zp\ boson production and decay
is characterised by three parameters: two effective coupling constants,
\gbl\ and \gy, and the \zp\ boson mass.
This parameterisation encompasses \zp\ bosons from many models, including the \zpchi\
belonging to the \esix-motivated model of the previous section, the \zpthreeR\ in a
left-right symmetric model~\cite{Senjanovic:1975rk,Mohapatra:1974hk} and the \zpBL\ of
the pure ($\mathrm{B-L}$) model~\cite{Basso:2008iv}. The minimal models are therefore particularly interesting for their generality, and because couplings are being directly constrained by the search.
The coupling parameter \gbl\ defines the coupling of a new \zp\ boson to the ($\mathrm{B-L}$)
current, while the \gy\ parameter represents the coupling to the weak hypercharge Y.
It is convenient to refer to the ratios $\gbltilde \equiv \gbl / g_Z$ and $\gytilde \equiv \gy / g_Z$,
where $g_Z$ is related to the coupling of the SM $Z$ boson to fermions defined by $g_Z = 2 M_Z / v$.
Here $v = 246$~\GeV{} is the SM Higgs vacuum expectation value.
To simplify further, the additional parameters \gammap~and~\thetamin\ are chosen as independent parameters with
the following definitions: $\gbltilde=\gammap \cos\thetamin$, $\gytilde=\gammap \sin\thetamin$.
The \gammap\ parameter measures the strength of the \zp\ boson coupling relative
to that of the SM $Z$ boson, while \thetamin\ determines the mixing between
the generators of the ($\mathrm{B-L}$) and weak hypercharge Y gauge groups.
Specific values of \gammap\ and \thetamin\ correspond to \zp\ bosons in various models,
as is shown in Table~\ref{tab:minimal_models} for the three cases mentioned in this section.

\begin{table}[h]
\caption{
Values for \gammap\ and \thetamin\ in the \MM\ corresponding to three specific \zp\ bosons: \zpBL, \zpchi\ and \zpthreeR.
The SM weak mixing angle is denoted by $\theta_{\mathrm{W}}$.
}
\label{tab:minimal_models}
\begin{center}
\begin{tabular}{l|ccc}
\hline
\hline
 & \zpBL & \zpchi & \zpthreeR  \\
\hline
\gammap & $\sqrt{\frac{5}{8}} \sin{\theta_{\mathrm{W}}}$ & $\sqrt{\frac{41}{24}} \sin{\theta_{\mathrm{W}}}$ & $\sqrt{\frac{5}{12}} \sin{\theta_W}$ \\
$\cos\thetamin$ & 1 & $\sqrt{\frac{25}{41}}$ & $\frac{1}{\sqrt{5}}$ \\
$\sin\thetamin$ & 0 & $-\sqrt{\frac{16}{41}}$ & $-\frac{2}{\sqrt{5}}$ \\
\hline
\hline
\end{tabular}
\end{center}
\end{table}

For the \MM, the width depends on \gammap\ and \thetamin, and the \zp\ interferes
with the SM \dy\ process. For example, taking the $\mathrm{3R}$ and $\mathrm{B-L}$ models investigated in this search, the width varies from less than 1\% up to 12.8\% and 39.5\% respectively, for the \gammap\ range considered. The branching fraction to leptons is the same as for the other \zp\ models considered in this search.
Couplings to hypothetical right-handed neutrinos, the Higgs boson, and to $W$~boson pairs are not considered.
Previous limits on the \zp\ mass versus \gammap\ were set by the \mbox{ATLAS} experiment.
For $\gammap=0.2$, the range of \zp\ mass limits at 95\%~CL corresponding to
$\thetamin \in [0,\pi]$ is 1.11~\TeV\ to 2.10~\TeV~\cite{Aad:2014cka}.

\subsection{Contact interactions}
\label{theory:CI}

Some models of physics beyond the SM result in non-resonant deviations
from the predicted SM dilepton mass spectrum.
Compositeness models motivated by the repeated pattern of quark and lepton generations
predict new interactions involving their constituents.
These interactions may be represented as a contact interaction
between initial-state quarks and final-state leptons~\cite{Eichten:1983hw, Eichten:1984eu}.
Other models producing non-resonant effects are models with large extra
dimensions~\cite{ArkaniHamed:1998nn} motivated by the hierarchy problem.
This search is sensitive to non-resonant new physics in these scenarios; however, constraints on these models are not evaluated in this article.

The following four-fermion CI Lagrangian~\cite{Eichten:1983hw, Eichten:1984eu}
is used to describe a new interaction in the process
$q\overline{q} \to \ell^+\ell^-$:

\vspace{-0.4cm}

\begin{eqnarray*}\label{lagrangian}
\mathcal L & = & \frac{g^2}{\Lambda^2}\;[ \eta_{\textrm{LL}} \, (\overline q_{\textrm{L}}\gamma_{\mu} q_{\textrm{L}})\,(\overline\ell_{\textrm{L}}\gamma^{\mu}\ell_{\textrm{L}})
 +\eta_{\textrm{RR}} (\overline q_{\textrm{R}}\gamma_{\mu} q_{\textrm{R}}) \,(\overline\ell_{\textrm{R}}\gamma^{\mu}\ell_{\textrm{R}}) \\ \nonumber
 &  & ~~~~~~~~ +\eta_{\textrm{LR}} (\overline q_{\textrm{L}}\gamma_{\mu} q_{\textrm{L}}) \,(\overline\ell_{\textrm{R}}\gamma^{\mu}\ell_{\textrm{R}})
+\eta_{\textrm{RL}} (\overline q_{\textrm{R}}\gamma_{\mu} q_{\textrm{R}}) \,(\overline\ell_{\textrm{L}}\gamma^{\mu}\ell_{\textrm{L}}) ] \: ,
\end{eqnarray*}

\noindent where $g$ is a coupling constant set to be $\sqrt{4\pi}$ by convention,
$\Lambda$ is the CI scale,
and $q_{\textrm{L,R}}$ and $\ell_{\textrm{L,R}}$ are left-handed and right-handed quark and lepton fields, respectively.
The symbol $\gamma_{\mu}$ denotes the gamma matrices, and the parameters $\eta_{ij}$, where $i$ and $j$ are L or R (left or right),
define the chiral structure of the new interaction.
Different chiral structures are investigated here, with the left--right (right--left) model obtained by setting $\eta_{\textrm{LR}} = \pm 1$ ($\eta_{\textrm{RL}} = \pm 1$) and all other parameters to zero. Likewise, the left--left and right--right models are obtained by setting the corresponding parameters
to $\pm 1$, and the others to zero. The sign of $\eta_{ij}$ determines whether the interference
between the SM Drell--Yan (DY) $q\overline{q} \to Z/\gamma^\ast \to \ell^+\ell^-$
process and the CI process is constructive ($\eta_{ij} = -1$) or destructive ($\eta_{ij} = +1$).

\section{ATLAS detector}
\label{sec:detector}
\newcommand{\AtlasCoordFootnote}{%
ATLAS uses a right-handed coordinate system with its origin at the nominal interaction point (IP)
in the centre of the detector and the $z$-axis along the beam pipe.
The $x$-axis points from the IP to the centre of the LHC ring,
and the $y$-axis points upwards.
Cylindrical coordinates $(r,\phi)$ are used in the transverse plane, 
$\phi$ being the azimuthal angle around the $z$-axis.
The pseudorapidity is defined in terms of the polar angle $\theta$ as $\eta = -\ln \tan(\theta/2)$.
Angular distance is measured in units of $\Delta R \equiv \sqrt{(\Delta\eta)^{2} + (\Delta\phi)^{2}}$.}

The ATLAS experiment~\cite{PERF-2007-01, IBL_TDR} at the LHC is a multipurpose particle detector
with a forward-backward symmetric cylindrical geometry and a near $4\pi$ coverage in 
solid angle.\footnote{\AtlasCoordFootnote}
It consists of an inner detector for tracking surrounded by a thin superconducting solenoid
providing a \SI{2}{\tesla} axial magnetic field, electromagnetic and hadronic calorimeters, and a muon spectrometer.
The inner detector (ID) covers the pseudorapidity range $|\eta| < 2.5$.
It consists of silicon pixel, silicon microstrip, and transition-radiation tracking detectors.
Lead/liquid-argon (LAr) sampling calorimeters provide electromagnetic (EM) energy measurements
with high granularity.
A hadronic (steel/scintillator-tile) calorimeter covers the central pseudorapidity range ($|\eta| < 1.7$).
The endcap and forward regions are instrumented with LAr calorimeters
for EM and hadronic energy measurements up to $|\eta| = 4.9$. The total thickness of the EM calorimeter is more than twenty radiation lengths.
The muon spectrometer (MS) surrounds the calorimeters and is based on
three large superconducting air-core toroids with eight coils each. The field integral of the toroids ranges between \num{2.0} and \SI{6.0}{\tesla\metre} for most of the detector.
The MS includes a system of precision tracking chambers and fast detectors for triggering.
A dedicated trigger system is used to select events.
The first-level trigger is implemented in hardware and uses the calorimeter and muon detectors to reduce the accepted rate to below \SI{100}{\kilo\hertz}. This is followed by a software-based trigger that reduces the accepted event rate to \SI{1}{\kilo\hertz} on average~\cite{Aaboud:2016leb}.

\section{Data and Monte Carlo samples}
\label{sec:data-mc}
This analysis uses data collected at the LHC during 2015 and 2016 $pp$ collisions at $\sqrt{s}$ = 13~\TeV{}. The total integrated luminosity corresponds to \lumill, considering the periods of data-taking with all sub-detectors functioning nominally. The event quality is also checked to remove events which contain noise bursts or coherent noise in the calorimeters.

Modelling of the various background sources primarily relies on MC simulation. The dominant background contribution arises from the DY process, which was simulated using the next-to-leading-order (NLO) \powhegbox~\cite{Alioli:2010xd} event generator, implementing the CT10~\cite{Lai:2010vv} parton distribution function (PDF), in conjunction with \pythiaeight.186~\cite{Sjostrand:2007gs} for event showering, and the ATLAS AZNLO set of tuned parameters~\cite{AZNLO:2014}. A more detailed description of this process is provided in Ref.~\cite{ATL-PHYS-PUB-2016-003}. The DY event yields are corrected with a rescaling that depends on the dilepton invariant mass from NLO to next-to-next-to-leading order (NNLO) in the strong coupling constant, computed with \vrap~0.9~\cite{Anastasiou:2003ds} and the CT14NNLO PDF set~\cite{Dulat:2015mca}. The NNLO quantum chromodynamic (QCD) corrections are a factor of $\sim$ 0.98 at a dilepton invariant mass ($m_{\ell\ell}$) of 3~\TeV{}. Mass-dependent electroweak (EW) corrections were computed at NLO with \mcsanc~1.20~\cite{Bondarenko:2013nu}. The NLO EW corrections are a factor of $\sim$ 0.86 at $m_{\ell\ell}$ = 3~\TeV{}. Those include photon-induced contributions ($\gamma\gamma \to \ell\ell$ via $t$- and $u$-channel processes) computed with the MRST2004QED PDF set~\cite{Martin:2004dh}.

Other backgrounds originate from top-quark~\cite{ATL-PHYS-PUB-2016-020} and diboson ($WW$, $WZ$, $ZZ$)~\cite{ATL-PHYS-PUB-2016-002} production. The diboson processes were simulated using \sherpa\ 2.2.1~\cite{Gleisberg:2008ta} with the CT10 PDF. The \ttbar\ and single-top-quark MC samples were simulated using the \powhegbox\ generator with the CT10 PDF, and are normalised to a cross-section as calculated with the Top++ 2.0 program~\cite{Czakon:2011xx}, which is accurate to NNLO in perturbative QCD, including resummation of next-to-next-to-leading logarithmic soft gluon terms. Background processes involving $W$ and $Z$ bosons decaying into $\tau$ lepton(s) were found to have a negligible contribution, and are not included. In the case of the dielectron channel, multi-jet and $W+$jets processes (which contribute due to the misidentification of jets as electrons) are estimated using a data-driven method, described in \refS{sec:background}.

Signal processes were produced at leading-order (LO) using \pythiaeight.186 with the NNPDF23LO PDF set~\cite{Ball:2012cx} and the ATLAS A14 set of tuned parameters~\cite{ATL-PHYS-PUB-2014-021} for event generation, parton showering and hadronisation. Interference effects (with DY production) are not included for the SSM and \esix\ model \zp\ signal due to large model dependence, but are included for the CI signal and for the Minimal model approach. Higher-order QCD corrections for the signal were computed with the same methodology as for the DY background. EW corrections were not applied to the \zp\ signal samples also due to the large model dependence. However, the EW corrections are applied to the CI signal samples, because interference effects are included.

The detector response is simulated with \geant~4~\cite{Agostinelli:2002hh}, and the events are processed with the same reconstruction software~\cite{Aad:2010ah} as used for the data. Furthermore, the distribution of the number of additional simulated $pp$ collisions in the same or neighbouring beam crossings (pile-up) is accounted for by overlaying minimum-bias events simulated with \pythiaeight.186 using the ATLAS A2 set of tuned parameters~\cite{ATL-PHYS-PUB-2014-021} and the MSTW2008LO PDF set~\cite{Martin:2009iq}, reweighting the MC simulation to match the distribution observed in the data.

\section{Event selection}
\label{sec:selection}
Dilepton candidates are selected in the data and simulated events by requiring at least one pair of reconstructed same-flavour lepton candidates (electrons or muons) and at least one reconstructed $pp$ interaction vertex, with the primary vertex defined as the one with the highest sum of track transverse momenta ($\pt$) squared.

Electron candidates are identified in the central region of the ATLAS detector
($|\eta|<\mathrm{2.47}$) by combining calorimetric and tracking information in a likelihood discriminant with four operating points: \textit{Very Loose}, \textit{Loose}, \textit{Medium} and \textit{Tight} each with progressively higher threshold for the discriminant, and stronger background rejection, as described in Ref.~\cite{Aaboud:2016vfy}.
The transition region between the central and forward regions of the calorimeters, in the range 1.37 $\le$ |$\eta$| $\le$ 1.52, exhibits poorer energy resolution and is therefore excluded. Electron candidates are required to have transverse energy (\et) greater than 30~\GeV{},
and a track consistent with the primary vertex both along the beamline
and in the transverse plane. The \textit{Medium} working point of the likelihood discrimination is used to select electron candidates while the \textit{Very Loose} and \textit{Loose} working points are used in the data-driven background estimation described in \refS{sec:background}.
In addition to the likelihood discriminant, selection criteria based
on track quality are applied.
The selection efficiency is approximately 96\% for electrons with \et\
between 30~\GeV{} and 500~\GeV{}, and decreases to approximately 95\%
for electrons with \et\ = 1.5 \TeV{}. The selection efficiency is evaluated in the data using a tag-and-probe
method~\cite{ATLAS-CONF-2016-024} up to \et\ of 500~\GeV{} and the uncertainties due to the modelling of the shower shape variables are estimated for electrons with higher \et\ using MC events, as described in Section~\ref{sec:systematics}.
The electron energy scale and resolution have been calibrated up to \et\ of 1~\TeV{}
using data collected at $\sqrt{s}$ = 8~\TeV{}~ and $\sqrt{s}$ = 13~\TeV{}~\cite{ATL-PHYS-PUB-2016-015}.
The energy resolution extrapolated for high-\et\ electrons (greater than 1~\TeV{}) is approximately 1\%.

Muon candidate tracks are, at first, reconstructed independently in the ID and the MS~\cite{MuonPerf}. The two tracks are then used as input to a combined fit (for \pt less than 300~\GeV{}) or to a statistical combination (for \pt greater than 300~\GeV{}). The combined fit takes into account the energy loss in the calorimeter and multiple-scattering effects. The statistical combination for high transverse momenta is performed to mitigate the effects of relative ID and MS misalignments.

In order to optimise momentum resolution, muon tracks are required
to have at least three hits in each of three precision chambers in the MS and
not to traverse regions of the MS which are poorly aligned.
This requirement reduces the muon reconstruction efficiency by about 20\% for
muons with a \pt\ greater than 1.5~\TeV{}.
Furthermore, muon candidates in the overlap of the MS barrel and endcap region ($1.01<|\eta|<1.10$) are rejected due to the potential relative misalignment between barrel and endcap.
Measurements of the ratio of charge to momentum (\qp) performed independently
in the ID and MS must agree within seven standard deviations, calculated
from the sum in quadrature of the ID and MS momentum uncertainties.
Finally, in order to reject events that contain a muon with poor track resolution in the MS, due to a low magnetic field integral and other effects, an event veto based on the MS track momentum measurement uncertainty is also applied. Muons are required to have \pt greater than 30~\GeV{}, $|\eta|<2.5$, and to be consistent with the primary vertex both along the beamline and in the transverse plane.

To further suppress background from misidentified jets as well as from light-flavour and
heavy-flavour hadron decays inside jets, lepton candidates are required to satisfy
calorimeter-based (only for electrons) and track-based (for both electrons and muons)
isolation criteria.
The calorimeter-based isolation relies on the ratio of the \et\
deposited in a cone of size $\Delta R = 0.2$, centered at the electron cluster barycentre,
to the total \et\ measured for the electron. The track-based isolation relies on the ratio of the summed scalar \pt\ of tracks within a variable-cone of size $\Delta R = 10\, \GeV/\pt$ to the \pt\ of the track associated with the candidate lepton. This variable-cone has no minimum size, meaning that the track-based isolation requirement effectively vanishes at very high lepton \pt.
The tracks are required to have $\pt > 1\,\GeV$, $|\eta| < 2.5$, meet all track quality criteria, and originate from the primary vertex. In all cases the contribution to the \et\ or \pt\ ascribed to the lepton candidate is removed from the isolation cone. The isolation criteria, applied to both leptons, have a fixed efficiency of 99\% over the full range of lepton momenta.

Calibration corrections are applied to electron (muon) candidates to match
energy (momentum) scale and resolution between data and simulation~\cite{Aad:2014nim, MuonPerf}.

Triggers were chosen to maximise the overall signal efficiency. In the dielectron channel, a two-electron trigger based on the \textit{Loose} identification
criteria with an \et\ threshold of 17~\GeV{} for each electron is used.
Events in the dimuon channel are required to pass at least one of two single-muon triggers
with \pt{} thresholds of 26~\GeV{} and 50~\GeV{}, with the former also requiring the muon
to be isolated.
These triggers select events from a simulated sample of \zpchi\ bosons with a pole mass of 3~\TeV{}
with an efficiency of approximately 86\% and 91\% for the dielectron and dimuon channels, respectively.

Data-derived corrections are applied in the samples to match the trigger,
reconstruction and isolation efficiencies between data and MC simulation. For each event with at least two same-flavour leptons, the dilepton candidate is built. If more than two electrons (muons) are found, the ones with the highest \et\ (\pt) are chosen. In the muon channel, only opposite-charge candidates are retained.
This requirement is not applied in the electron channel due to a higher chance
of charge misidentification for high-\et\ electrons.
There is no explicit overlap removal between the dielectron and dimuon channel, but a negligible number of common events at low dilepton masses enter the combination.

Representative values of the total acceptance times efficiency for a \zpchi\ boson with a pole mass of 3~\TeV{} are 71\% in the dielectron channel and 40\% in the dimuon channel.

\section{Background estimation}
\label{sec:background}
The backgrounds from processes including two real leptons in the final state (DY, $t\bar{t}$, single top quark, $WW$, $WZ$, and $ZZ$ production) are modelled using the MC samples described in \refS{sec:data-mc}. In the mass range 120~\GeV{} < $\mll$ < 1~\TeV{} the corrected DY background is smoothed to remove statistical fluctuations due to the limited MC sample size compared to the large integrated luminosity of the data, by fitting the spectrum and using the resulting fitted function to set the expected event yields in that mass region. The chosen fit function consists of a relativistic Breit--Wigner function with mean and width fixed to $M_{Z}$ and $\Gamma_Z$ respectively~\cite{Olive:2016xmw}, multiplied by an analytic function taking into account detector resolution, selection efficiency, parton distribution function effects, and contributions from the photon-induced process and virtual photons. At higher dilepton invariant masses the statistical uncertainty of the MC simulation is much smaller than that of the data through the use of mass-binned MC samples.

An additional background arises from $W+$jets and multi-jet events from which at most one real lepton is produced. This background contributes to the selected samples due to having one or more jets satisfying the lepton selection criteria (so called ``fakes''). In the dimuon channel, contributions from $W+$jets and multi-jet production are found to be negligible, and therefore are not included in the expected yield. In the dielectron channel the contributions from these processes are determined with a data-driven technique, the \textit{matrix method}, in two steps. In the first step, the probabilities that a jet and a real electron satisfy the electron identification requirements are evaluated, for both the nominal and a loosened selection criteria. The loosened selection differs from the nominal one by the use of the \textit{Loose} electron identification criteria and no isolation criterion. Then, in the second step these probabilities are used to estimate the level of contamination, due to fakes, in the selected sample of events.

A probability $r$ that a real electron passing the loosened selection satisfies the nominal electron selection criteria is estimated from MC simulated DY samples in several regions of \et\ and $|\myeta|$. The probability $f$ that a jet passing the loosened selection satisfies the nominal electron selection criteria is determined in regions of \et\ and $|\myeta|$ in data samples triggered on the presence of a $\textit{Very Loose}$ or a $\textit{Loose}$ electron candidate. Contributions to these samples from the production of $W$ and $Z$ bosons are suppressed by vetoing events with large missing transverse energy ($\met > 25$ \GeV) or with two $\textit{Loose}$ electron candidates compatible with $Z$ boson mass, or two candidates passing the \textit{Medium} identification criteria. The \met\ is reconstructed as the negative vectorial sum of the calibrated momenta of the electrons, muons, and jets, in the event. Residual contributions from processes with real electrons in the calculation of $f$ are accounted for by using the MC simulated samples.

The selected events are grouped according to the identification criteria satisfied by the electrons. A system of equations between numbers of paired objects ($N_{ab}$, with $\et^{a} > \et^{b}$) is used to solve for the unknown contribution to the background in each of the kinematic regions from events with one or more fake electrons:
\begin{equation}
\label{Eq:MM:1}
\begin{pmatrix}
N_{\mathrm{TT}}\\
N_{\mathrm{TL}}\\
N_{\mathrm{LT}}\\
N_{\mathrm{LL}}
\end{pmatrix}
=
\begin{pmatrix}
r^2 & rf & fr & f^2  \\
r(1-r) & r (1-f) & f(1-r) & f(1-f) \\
(1-r)r & (1-r)f & (1-f)r & (1-f)f \\
(1-r)^2 & (1-r)(1-f) & (1-f)(1-r) & (1-f)^2
\end{pmatrix}
\begin{pmatrix}
N_{\mathrm{RR}}\\
N_{\mathrm{RF}}\\
N_{\mathrm{FR}}\\
N_{\mathrm{FF}}
\end{pmatrix}
\end{equation}
Here the subscripts $\mathrm{R}$ and $\mathrm{F}$ refer to real electrons and fakes (jets), respectively. The subscript $\mathrm{T}$ refers to electrons that satisfy the nominal selection criteria. The subscript $\mathrm{L}$ corresponds
to electrons that pass the loosened requirements described above but fail the nominal requirements.

The background is given as the part of $N_{\mathrm{TT}}$ that originates from a pair of objects with at least one fake electron:
\begin{eqnarray}
\label{Eq:MM:2}
N^{\text{Multi-jet\ \&\ W+jets}} =  rf (N_{\mathrm{RF}}+ N_{\mathrm{FR}}) + f^2N_{\mathrm{FF}}
\end{eqnarray}
The true paired objects on the right-hand side of Eq.~(\ref{Eq:MM:2}) can be expressed in terms of measureable quantities ($N_{\mathrm{TT}}, N_{\mathrm{TL}}, N_{\mathrm{LT}}, N_{\mathrm{LL}})$ by inverting the matrix in Eq.~(\ref{Eq:MM:1}).

The estimate is extrapolated to the full mass range considered by fitting an analytic function to the dielectron invariant mass (\mee{}) distribution above $\sim$125~\GeV{} to mitigate effects of limited event counts in the high-mass region and of method instabilities due to a negligible contribution from fakes in the $Z$ peak region. The fit is repeated by increasing progressively the lower edge of the fit range by $\sim$10 \GeV\ per step until $\sim$195~\GeV. The weighted mean of all fits is taken as the central value and the envelope as the uncertainty. Additional uncertainties in this background estimate are evaluated by considering differences between the estimates for events with same-charge and opposite-charge electrons as well as by varying the electron identification probabilities. The uncertainty on this background can, due to the extrapolation, become very large at high mass, but has only a negligible impact on the final results of this analysis.

\section{Systematic uncertainties}
\label{sec:systematics}
Systematic uncertainties estimated to have a non-negligible impact on the expected cross-section limit are considered as nuisance parameters in the statistical interpretation and include both the theoretical and experimental effects on the total background and experimental effects on the signal.

Theoretical uncertainties in the background prediction are dominated by the DY background, throughout the entire dilepton invariant mass range. They arise from the eigenvector variations of the nominal PDF set, as well as variations of PDF scales, the strong coupling (\alphas(M$_{Z}$)), EW corrections, and photon-induced (PI) corrections. The effect of choosing different PDF sets are also considered. The theoretical uncertainties are the same for both channels at generator level, but they result in different uncertainties at reconstruction level due to the differing resolutions between the dielectron and dimuon channels.

The PDF variation uncertainty is obtained using the 90\% CL CT14NNLO PDF error set and by following the procedure described in Refs.~\cite{Aad:2014cka, Gao:2013bia, Butterworth:2015oua}. Rather than using a single nuisance parameter to describe the 28 eigenvectors of this PDF error set, which could lead to an underestimation of its effect, a re-diagonalised set of 7 PDF eigenvectors was used~\cite{Dulat:2015mca}, which are treated as separate nuisance parameters. This represents the minimal set of PDF eigenvectors that maintains the necessary correlations, and the sum in quadrature of these eigenvectors matches the original CT14NNLO error envelope well. The uncertainties due to the variation of PDF scales and \alphas\ are derived using \vrap\ with the former obtained by varying the renormalisation and factorisation scales of the nominal CT14NNLO PDF up and down simultaneously by a factor of two. The value of \alphas\ used (0.118) is varied by $\pm 0.003$. The EW correction uncertainty was assessed by comparing the nominal additive (1+$\delta_{\mathrm{EW}}$+$\delta_{\mathrm{QCD}}$) treatment with the multiplicative approximation ((1+$\delta_{\mathrm{EW}}$)(1+$\delta_{\mathrm{QCD}}$)) treatment of the EW correction in the combination of the higher-order EW and QCD effects. The uncertainty in the photon-induced correction is calculated based on the uncertainties in the quark masses and the photon PDF. Following the recommendations of the PDF4LHC forum~\cite{Butterworth:2015oua}, an additional uncertainty due to the choice of nominal PDF set is derived by comparing the central values of CT14NNLO with those from other PDF sets, namely MMHT14~\cite{Motylinski:2014sya} and NNPDF3.0~\cite{Ball:2014uwa}. 
The maximum absolute deviation from the envelope of these comparisons is used as the PDF choice uncertainty, where it is larger than the CT14NNLO PDF eigenvector variation envelope. Theoretical uncertainties are not applied to the signal prediction in the statistical interpretation.

Theoretical uncertainties on the estimation of the top quark and diboson backgrounds were also considered, both from the independent variation of the factorisation ($\mu_{\text{F}}$) and renormalisation ($\mu_{\text{R}}$) scales, and from the variations in the PDF and \alphas, following the PDF4LHC prescription. Normalisation uncertainties in the top quark and diboson background are shown in the ``Top Quarks Theoretical'' and ``Dibosons Theoretical'' entry in Table~\ref{systematics}.

\begin{table}
  \begin{center}
    \caption{Summary of the pre-marginalised relative systematic uncertainties in the expected number of events at dilepton masses of 2~\TeV{} and 4~\TeV{}. The values reported in parenthesis correspond to the 4~\TeV{} mass. The values quoted for the background represent the relative change in the total expected number of events in the corresponding \mll\ histogram bin containing the reconstructed \mll\ mass of 2~\TeV~(4~\TeV). For the signal uncertainties the values were computed using a \zpchi\ signal model with a pole mass of 2~\TeV~(4~\TeV) by comparing yields in the core of the mass peak (within the full width at half maximum) between the distribution varied due to a given uncertainty and the nominal distribution. ``-'' represents cases where the uncertainty is not applicable. \label{systematics}}
    \centering
     \begin{tabular}{l|cc|cc}
      \toprule
      \toprule
      Source  &  \multicolumn{2}{c|}{Dielectron channel [\%]}  &  \multicolumn{2}{c}{Dimuon channel [\%]}  \\
      &  Signal  &  Background  &  Signal  &  Background  \\
      \midrule
Luminosity & \syspair{3.2}{3.2} & \syspairb{3.2}{3.2} & \syspair{3.2}{3.2} & \syspair{3.2}{3.2} \\
MC statistical & \syspair{\lessthan{1.0}}{\lessthan{1.0}} & \syspairb{\lessthan{1.0}}{\lessthan{1.0}} & \syspair{\lessthan{1.0}}{\lessthan{1.0}} & \syspair{\lessthan{1.0}}{\lessthan{1.0}} \\
Beam energy & \syspair{2.0}{4.1} & \syspairb{2.0}{4.1} & \syspair{1.9}{3.1} & \syspair{1.9}{3.1} \\
Pile-up effects & \syspair{\lessthan{1.0}}{\lessthan{1.0}} & \syspairb{\lessthan{1.0}}{\lessthan{1.0}} & \syspair{\lessthan{1.0}}{\lessthan{1.0}} & \syspair{\lessthan{1.0}}{\lessthan{1.0}} \\
\hline
DY PDF choice & - & \syspairb{\lessthan{1.0}}{8.4} & - & \syspair{\lessthan{1.0}}{1.9} \\
DY PDF variation & - & \syspairb{8.7}{19} & - & \syspair{7.7}{13} \\
DY PDF scales & - & \syspairb{1.0}{2.0} & - & \syspair{\lessthan{1.0}}{1.5} \\
DY \alphas\ & - & \syspairb{1.6}{2.7} & - & \syspair{1.4}{2.2} \\
DY EW corrections & - & \syspairb{2.4}{5.5} & - & \syspair{2.1}{3.9} \\
DY $\gamma$-induced corrections & - & \syspairb{3.4}{7.6} & - & \syspair{3.0}{5.4} \\
\hline
Top quarks theoretical & - & \syspairb{\lessthan{1.0}}{\lessthan{1.0}} & - & \syspair{\lessthan{1.0}}{\lessthan{1.0}} \\
Dibosons theoretical & - & \syspairb{\lessthan{1.0}}{\lessthan{1.0}} & - & \syspair{\lessthan{1.0}}{\lessthan{1.0}} \\
\hline
Reconstruction efficiency & \syspair{\lessthan{1.0}}{\lessthan{1.0}} & \syspairb{\lessthan{1.0}}{\lessthan{1.0}} & \syspair{10}{17} & \syspair{10}{17} \\
Isolation efficiency & \syspair{9.1}{9.7} & \syspairb{9.1}{9.7} & \syspair{1.8}{2.0} & \syspair{1.8}{2.0} \\
Trigger efficiency & \syspair{\lessthan{1.0}}{\lessthan{1.0}} & \syspairb{\lessthan{1.0}}{\lessthan{1.0}} & \syspair{\lessthan{1.0}}{\lessthan{1.0}} & \syspair{\lessthan{1.0}}{\lessthan{1.0}} \\
Identification efficiency & \syspair{2.6}{2.4} & \syspairb{2.6}{2.4} & - & - \\
\hline
Lepton energy scale & \syspair{\lessthan{1.0}}{\lessthan{1.0}} & \syspairb{4.1}{6.1} & \syspair{\lessthan{1.0}}{\lessthan{1.0}} & \syspair{\lessthan{1.0}}{\lessthan{1.0}} \\
Lepton energy resolution & \syspair{\lessthan{1.0}}{\lessthan{1.0}} & \syspairb{\lessthan{1.0}}{\lessthan{1.0}} & \syspair{2.7}{2.7} & \syspair{\lessthan{1.0}}{6.7} \\
\hline
Multi-jet \& $W$+jets & - & \syspairb{10}{129} & - & - \\
\midrule
Total & \syspair{10}{11} & \syspairb{18}{132} & \syspair{11}{18} & \syspair{14}{24} \\
      \bottomrule
      \bottomrule
    \end{tabular}
  \end{center}
\end{table}

The following sources of experimental uncertainty are accounted for: lepton efficiencies due to triggering, identification, reconstruction, and isolation, lepton energy scale and resolution, pile-up effects, as well as the multi-jet and $W$+jets background estimate. The same sources of experimental uncertainty are considered for the DY background and signal treatment. Efficiencies are evaluated using events from the $Z \to \ell\ell$ peak and then extrapolated to high energies. The uncertainty in the muon reconstruction efficiency is the largest experimental uncertainty in the dimuon channel. It includes the uncertainty obtained from $Z \to \mu\mu$ data studies and a high-\pt\ extrapolation uncertainty corresponding to the decrease in the muon reconstruction and selection efficiency with increasing \pt\ which is predicted by the MC simulation. The effect on the muon reconstruction efficiency was found to be approximately 3\% per \TeV{} as a function of muon \pt. The uncertainty in the electron identification efficiency extrapolation is based on the differences in the electron shower shapes in the EM calorimeters between data and MC simulation in the $Z \to ee$ peak, which are propagated to the high-\et\ electron sample. The effect on the electron identification efficiency was found to be 2.0\% and is independent of \et\ for electrons with \et\ above 150~\GeV{}. For the isolation efficiencies, uncertainties were estimated for 150 $<$ \pt\ $<$ 500~\GeV{} and above 500~\GeV{} separately, using DY candidates in data. The larger isolation uncertainty that is observed for electrons is due to the uncertainty inherent in calorimeter-based isolation for electrons (track-based isolation is also included), compared to the solely track-based only isolation for muons. Mismodelling of the muon momentum resolution due to residual misalignments in the MS can alter the steeply falling background shape at high dilepton mass and can significantly modify the width of the signals line shape. This uncertainty is obtained by studying the muon momentum resolution in dedicated data-taking periods with no magnetic field in the MS~\cite{MuonPerf}. For the dielectron channel, the uncertainty includes a contribution from the multi-jet and $W$+jets data-driven estimate that is obtained by varying both the overall normalisation and the extrapolation methodology, which is explained in \refS{sec:background}. The systematic uncertainty from pile-up effects is assessed by inducing a variation in the pile-up reweighting of MC events and is included to cover the uncertainty on the ratio of the predicted and measured inelastic cross-section in the fiducial volume defined by M$_X$ $>$ 13~\GeV{}, where M$_X$ is the mass of the non-diffractive hadronic system~\cite{Aaboud:2016mmw}. An uncertainty on the beam energy of 0.65\% is estimated and included. The uncertainty on the combined 2015 and 2016 integrated luminosity is 3.2\%. It is derived, following a methodology similar to that detailed in Ref.~\cite{Aaboud:2016hhf}, from a calibration of the luminosity scale using $x$--$y$ beam-separation scans performed in August 2015 and May 2016. Systematic uncertainties used in the statistical analysis of the results are summarised in Table~\ref{systematics} at dilepton mass values of 2~\TeV{} and 4~\TeV{}. The systematic uncertainties are constrained in the likelihood during the statistical interpretation through a marginalisation procedure, as described in Section~\ref{sec:statistics}.

\section{Event yields}
\label{sec:yields}
Expected and observed event yields, in bins of invariant mass, are shown in~\refT{ee_yields_table} for the dielectron channel, and in~\refT{mm_yields_table} for the dimuon channel. Expected event yields are split into the different background sources and the yields for two signal scenarios are also provided.
In general, the observed data are in good agreement with the SM prediction, taking into account the uncertainties as described in \refS{sec:systematics}.

\begin{table}[!h]
\caption{Expected and observed event yields in the dielectron channel in different dilepton mass intervals. The quoted errors correspond to the combined statistical, theoretical, and experimental systematic uncertainties. Expected event yields are reported for the \zpchi\ model, for two values of the pole mass. All numbers shown are obtained before the marginalisation procedure.}
\label{ee_yields_table}
\centering
\vspace{0.2cm}
\footnotesize
\begin{tabular}{l r@{ $\pm$ }l r@{ $\pm$ }l r@{ $\pm$ }l r@{ $\pm$ }l r@{ $\pm$ }l }
\toprule
\toprule
\mee\ [\GeV]           & \multicolumn{2}{c}{80--120}   & \multicolumn{2}{c}{120--250} & \multicolumn{2}{c}{250--400} & \multicolumn{2}{c}{400--500} & \multicolumn{2}{c}{500--700} \\
\midrule
Drell--Yan             & 11\,800\,000   & 700\,000   & 216\,000    & 11\,000        & 17\,230     & 1000         & 2640   & 180               & 1620  & 120     \\
Top quarks             &  28\,600       & 1800       &  44\,600    &    2900        &    8300     &  600         & 1130   &  80               &  560   & 40     \\
Dibosons               &  31\,400       & 3300       &     7000    &     700        &    1300     &  140         &  228   &  25               &  146   & 16     \\
Multi-jet \& $W$+jets  &  11\,000       & 9000       &     5600    &    2000        &     780     &   80         &  151   &  21               &  113   & 17     \\
\midrule
Total SM               & 11\,900\,000   & 700\,000   & 273\,000    & 12\,000        & 27\,600     & 1100         & 4150   & 200               & 2440  & 130     \\
\midrule
Data                   & \multicolumn{2}{c}{ 12\,415\,434} & \multicolumn{2}{c}{ 275\,711}  & \multicolumn{2}{c}{ 27\,538}   & \multicolumn{2}{c}{ 4140}    & \multicolumn{2}{c}{ 2390} \\
\midrule
\zpchi\ (4~\TeV{})     & 0.00635    & 0.00021          & 0.0390    & 0.0015           & 0.0564    & 0.0025           & 0.0334 & 0.0027              & 0.064   & 0.004  \\
\zpchi\ (5~\TeV{})     & 0.00305    & 0.00012          & 0.0165    & 0.0006           & 0.0225    & 0.0010           & 0.0139 & 0.0007              & 0.0275  & 0.0015 \\
\bottomrule
\toprule
\mee\ [\GeV]           & \multicolumn{2}{c}{700--900} & \multicolumn{2}{c}{900--1200} & \multicolumn{2}{c}{1200--1800} & \multicolumn{2}{c}{1800--3000} & \multicolumn{2}{c}{3000--6000} \\
\midrule
Drell--Yan             & 421 & 34 & 176  & 17  & 62  & 7   & 8.7  & 1.3  & 0.34  & 0.07  \\
Top quarks             & 94  & 8  & 27.9 & 2.8 & 5.1 & 0.7 & \multicolumn{2}{c}{$<$ 0.001} & \multicolumn{2}{c}{$<$ 0.001} \\
Dibosons               & 39  & 4  & 16.9 & 2.1 & 5.8 & 0.8 & 0.74 & 0.11 & 0.028 & 0.004 \\
Multi-jet \& $W$+jets  & 39 & 6 & 16.1 & 2.0 & 7.9 & 2.3 & 1.6  & 1.2  & 0.08  & 0.27  \\
\midrule
Total SM               & 590 & 40 & 237 & 17 & 81 & 7 & 11.0 & 1.8 & 0.45 & 0.28 \\
\midrule
Data                   & \multicolumn{2}{c}{ 589} & \multicolumn{2}{c}{ 209} & \multicolumn{2}{c}{ 61} & \multicolumn{2}{c}{ 10} & \multicolumn{2}{c}{ 0 } \\
\midrule
\zpchi\ (4~\TeV{})     & 0.0585 & 0.0035 & 0.074  & 0.005  & 0.121 & 0.011 & 0.172 & 0.017 & 2.57  & 0.27   \\
\zpchi\ (5~\TeV{})     & 0.0218 & 0.0013 & 0.0295 & 0.0021 & 0.040 & 0.004 & 0.040 & 0.004 & 0.280 & 0.030  \\
\bottomrule
\bottomrule
\end{tabular}
\end{table}

\begin{table}[!h]
  \caption{Expected and observed event yields in the dimuon channel in different dilepton mass intervals. The quoted errors correspond to the combined statistical, theoretical, and experimental systematic uncertainties. Expected event yields are reported for the \zpchi\ model, for two values of the pole mass. All numbers shown are obtained before the marginalisation procedure.}
  \label{mm_yields_table}
  \centering
  \vspace{0.2cm}
  \footnotesize

  \begin{tabular}{l r@{ $\pm$ }l r@{ $\pm$ }l r@{ $\pm$ }l r@{ $\pm$ }l  r@{ $\pm$ }l }
    \toprule
    \toprule
    \mmm\ [\GeV]        & \multicolumn{2}{c}{80--120}  & \multicolumn{2}{c}{120--250} & \multicolumn{2}{c}{250--400} & \multicolumn{2}{c}{400--500} & \multicolumn{2}{c}{500--700} \\
    \midrule
    Drell--Yan          & 10\,700\,000   & 600\,000    & 177\,900  & 10\,000          & 12\,200  & 700               & 1770   & 120               & 1060 & 80  \\
    Top quarks          &      24\,700   &     1700    &  34\,200  &    2400          &    6100  & 500               &  830   & 70                &  401 & 33  \\
    Dibosons            &      26\,000   &     2800    &     5400  &     600          &     910  & 100               &  155   & 17                &   93 & 11  \\
    \midrule
    Total SM            & 10\,800\,000   & 600\,000    & 218\,000  & 10\,000          & 19\,200  & 900               & 2760   & 140               & 1550 & 90  \\
    \midrule
    Data                & \multicolumn{2}{c}{ 11\,321\,561} & \multicolumn{2}{c}{ 224\,703} & \multicolumn{2}{c}{ 19\,239}   & \multicolumn{2}{c}{ 2766}    & \multicolumn{2}{c}{ 1532}  \\
    \midrule
    \zpchi\ (4~\TeV{})  & 0.00873    & 0.00032         & 0.0334 & 0.0015              & 0.0441 & 0.0021              & 0.0246 & 0.0014              & 0.052  & 0.004   \\
    \zpchi\ (5~\TeV{})  & 0.00347    & 0.00014         & 0.0137 & 0.0006              & 0.0151 & 0.0007              & 0.0105 & 0.0006              & 0.0176 & 0.0012  \\
    \bottomrule
    \toprule
    \mmm\ [\GeV]        & \multicolumn{2}{c}{700--900} & \multicolumn{2}{c}{900--1200} & \multicolumn{2}{c}{1200--1800} & \multicolumn{2}{c}{1800--3000} & \multicolumn{2}{c}{3000--6000} \\
    \midrule
    Drell--Yan          & 263  & 23  & 110  & 11  & 37  & 4   & 5.4  & 0.8  & 0.30   & 0.07  \\
    Top quarks          & 68   & 6   & 24.5 & 3.0 & 5.3 & 0.9 & 0.11 & 0.08 & \multicolumn{2}{c}{$<$ 0.001} \\
    Dibosons            & 24.3 & 2.9 & 9.8  & 1.2 & 3.2 & 0.4 & 0.45 & 0.07 & 0.0184 & 0.0035 \\
    \midrule
    Total SM            & 355  & 24  & 144  & 11  & 45  & 4   & 6.0  & 0.8  & 0.32   & 0.07 \\
    \midrule
    Data                & \multicolumn{2}{c}{ 322} & \multicolumn{2}{c}{ 141} & \multicolumn{2}{c}{ 48} & \multicolumn{2}{c}{ 4} & \multicolumn{2}{c}{ 0 } \\
    \midrule
    \zpchi\ (4~\TeV{})  & 0.0362 & 0.0026 & 0.048  & 0.004  & 0.067  & 0.006  & 0.186  & 0.022  & 1.24  & 0.19   \\
    \zpchi\ (5~\TeV{})  & 0.0153 & 0.0011 & 0.0185 & 0.0015 & 0.0233 & 0.0021 & 0.0258 & 0.0029 & 0.118 & 0.020  \\
    \bottomrule
    \bottomrule
  \end{tabular}
\end{table}

Distributions of \mll{} in the dielectron and dimuon channels are shown in Figure~\ref{plots_inv_mass}. No clear excess is observed, but significances are quantified and discussed in Section~\ref{sec:statistics}. The highest dilepton invariant mass event is \HighestMee~\TeV{} in the dielectron channel, and \HighestMmumu~\TeV{} in the dimuon channel. Both of these events are well-measured with little other detector activity.

\begin{figure}[!htb]
  \subfloat[][]{
    \includegraphics[width=0.49\linewidth]{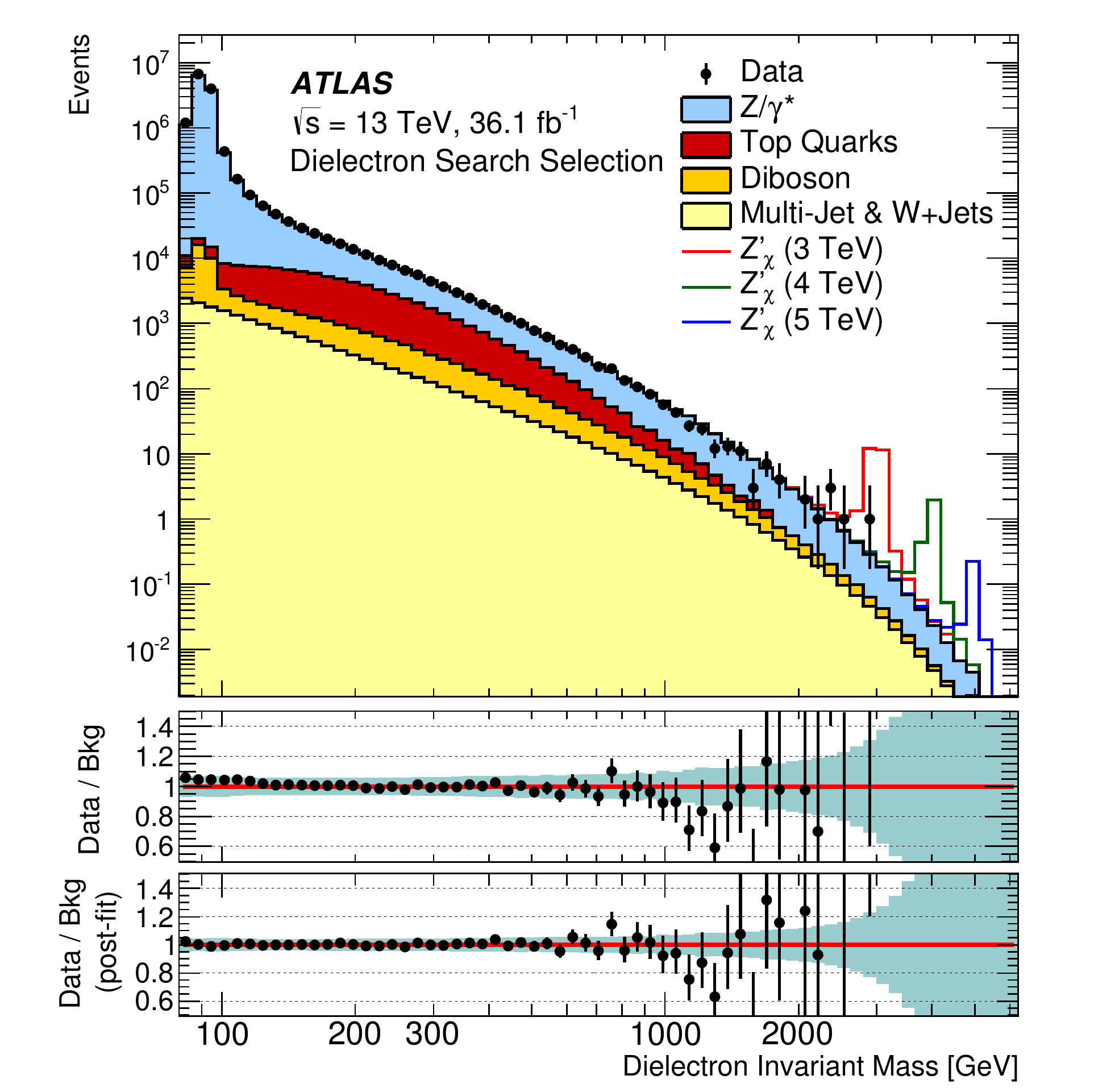}\label{plots_inv_mass_ee}
  }
  \subfloat[][]{
    \includegraphics[width=0.49\linewidth]{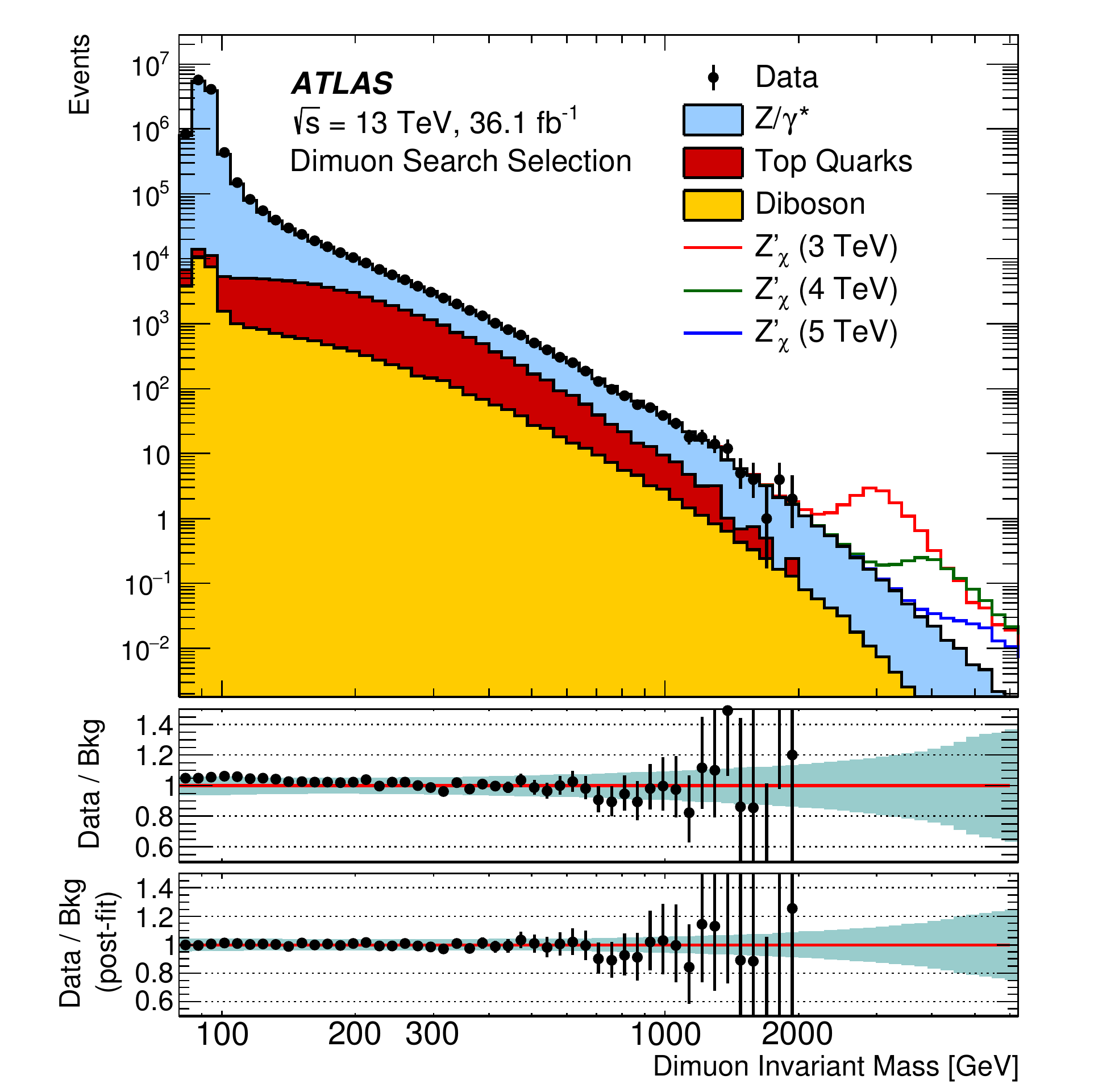}\label{plots_inv_mass_mm}
  }
  \caption{Distributions of \protect\subref{plots_inv_mass_ee} dielectron and \protect\subref{plots_inv_mass_mm} dimuon reconstructed invariant mass (\mll) after selection, for data and the SM background estimates as well as their ratio before and after marginalisation. Selected \zpchi\ signals with a pole mass of 3, 4 and 5~\TeV\ are overlaid. The bin width of the distributions is constant in log(\mll) and the shaded band in the lower panels illustrates the total systematic uncertainty, as explained in~\refS{sec:systematics}. The data points are shown together with their statistical uncertainty. Exact bin edges and contents are provided in \refT{ee_log_table} and \refT{mm_log_table} in the appendix.}
  \label{plots_inv_mass}
\end{figure}

\section{Statistical analysis}
\label{sec:statistics}
The \mll{} distributions are scrutinised for a resonant or non-resonant new physics excess using two methods and are used to set limits on resonant and non-resonant new physics models, as well as on generic resonances. Tabulated values of all the observed results, along with their uncertainties, are also provided in the Durham HEP database~\cite{HepData}.
The signal search and limit setting rely on a likelihood function, dependent on the parameter of interest, such as the signal cross-section, signal strength, coupling constant or the contact interaction scale. The likelihood function also depends on nuisance parameters which describe the systematic uncertainties. In this analysis the data are assumed to be Poisson-distributed in each bin of the \mll{} distribution and the likelihood is constructed as a product of individual bin likelihoods. In case of the individual channel results, the product is taken over the bins of the \mll{} histogram in the given channel, while for combined results the product is taken over bins of histograms in dielectron and dimuon channels. The logarithmic \mll\ histogram binning shown in Figure~\ref{plots_inv_mass} uses 66 mass bins and is chosen for setting limits on resonant signals. This binning is optimal for resonances with a width of 3\%, therefore the chosen bin width for the \mll{} histogram in the search phase corresponds to the resolution in the dielectron (dimuon) channel, which varies from 10 (60)~\GeV{} at \mll\ = 1~\TeV{} to 15 (200)~\GeV{} at \mll\ = 2~\TeV{}, and 20 (420)~\GeV{} at \mll\ = 3~\TeV{}. For setting limits on the contact interaction scale, the \mll{} distribution has eight bins above 400~\GeV{} with bin widths varying from 100 to 1500~\GeV{}. The \mll{} region from 80 to 120~\GeV{} is included in the likelihood as a single bin in the limit setting on resonant signals to help constrain mass-independent components of systematic uncertainties, but that region is not searched for a new-physics signal.

The parameter $\mu$ is defined as the ratio of the signal production cross-section times branching ratio into the dilepton final state (\xbr) to its theoretically predicted value. Upper limits on \xbr\ for specific \zp\ boson models and generic \zp\ bosons, $\gamma'$ of the Minimal \zp\ boson, and lower limit on the CI scale $\Lambda$ are set in a Bayesian approach. The calculations are performed with the Bayesian Analysis Toolkit (BAT)~\cite{Caldwell:2008fw}, which uses a Markov Chain MC (MCMC) technique to compute the marginal posterior probability density of the parameter of interest (so-called ``marginalisation''). Limit values obtained using the experimental data are quoted as observed limits, while median values of the limits obtained from a large number of simulated experiments, where only SM background is present, are quoted as the expected limits. The upper limits on \xbr\ are interpreted as lower limits on the \zp\ pole mass using the relationship between the pole mass and the theoretical \zp\ cross-section. In the context of the Minimal \zp\ model or CI scenarios, limits are set on the parameter of interest. In the case of the Minimal \zp\ model the parameter of interest is $\gamma'^4$. For a CI the parameter of interest is set either to $1/\Lambda^2$ or to $1/\Lambda^4$ as this corresponds to the scaling of the CI--SM interference contribution or the pure CI contribution respectively. In both the Minimal \zp\ and the CI cases, the nominal Poisson expectation in each \mll{} bin is expressed as a function of the parameter of interest. As in the context of the \zp\ limit setting, the Poisson mean is modified by shifts due to systematic uncertainties, but in both the Minimal \zp\ and the CI cases, these shifts are non-linear functions of the parameter of interest. A prior uniform in the parameter of interest is used for all limits.   

Two complementary approaches are used in the search for a new-physics signal. The first approach, which does not rely on a specific signal model and therefore is sensitive to a wide range of new physics, uses the {\textsc{BumpHunter}} (BH)~\cite{Choudalakis:2011qn} utility. In this approach, all consecutive intervals in the \mll{} histogram ranging from two bins to half of the bins in the histogram are searched for an excess. In each such interval a Poisson probability ($p$-value) is computed for an event count greater or equal to the number observed found in data, given the SM prediction.
The modes of marginalised posteriors of the nuisance parameters from the MCMC method are used to construct the SM prediction. The negative logarithm of the smallest $p$-value is the BH statistic. The BH statistic is then interpreted as a global $p$-value utilising simulated experiments where, in each simulated experiment, simulated data is generated from SM background model. The dielectron and dimuon channels are tested separately. 

A search for \zpchi\ signals as well as generic \zp\ signals with widths from 1\% to 12\% is performed utilising the log-likelihood ratio (LLR) test described in \citeR{stat_asymptotic}. This second approach is specifically sensitive to narrow \zp-like signals, and is thus complementary to the more general BH approach. To perform the LLR search, the Histfactory~\cite{Cranmer:1456844} package is used together with the RooStats~\cite{RooStats} and RooFit~\cite{Verkerke:2003ir} packages. The $p$-value for finding a \zpchi\ signal excess (at a given pole mass), or a variable width generic \zp\ excess (at a given central mass and with a given width), more significant than that observed in the data, is computed analytically, using a test statistic $q_0$. The test statistic $q_0$ is based on the logarithm of the profile likelihood ratio $\lambda(\mu)$. The test statistic is modified for signal masses below 1.5~\TeV{} to also quantify the significance of potential deficits in the data.
As in the BH search the SM background model is constructed using the modes of marginalised posteriors of the nuisance parameters from the MCMC method, and these nuisance parameters are not included in the likelihood at this stage. Therefore, in the search-phase the background estimate and signal shapes are fixed to their post-marginalisation estimates, and systematic uncertainties are not included in the computation of the $p$-value. Starting with \mzp{} = 150 \GeV{}, multiple mass hypotheses are tested in pole-mass steps corresponding to the histogram bin width to compute the local $p$-values --- i.e.\ $p$-values corresponding to specific signal mass hypotheses.  Simulated experiments (for \mzp{} $>$ 1.5~\TeV{}) and asymptotic relations (for \mzp{} $<$ 1.5~\TeV{}) in \citeR{stat_asymptotic} are used to estimate the global $p$-value, which is the probability to find anywhere in the \mll{} distribution a \zp-like excess more significant than that observed in the data.

\section{Results}
\label{sec:result}
The data, scrutinised using the statistical tests described in the previous section, show no significant excesses.
The LLR tests for a \zpchi\ resonance find global $p$-values of \LLRpvalee{}, \LLRpvalmm\, and \LLRpvalcomb\ in the dielectron, dimuon, and combined channels, respectively. The local and global $p$-values as a function of the \zp\ pole mass are shown in Figure~\ref{fig:observedSignificances}. The un-capped $p$-value, is used below a pole mass of 1.5~\TeV{}, which quantifies both excesses and deficits, while above 1.5~\TeV{} the signal strength parameter is constrained to be positive, yielding a capped p-value. This constraint is used in the high mass region where the expected background is very low, to avoid ill-defined configurations of the probability density function in the likelihood fit, with negative probabilities.

The largest deviation from the background-only hypothesis using the LLR tests for a \zpchi\ is observed at 2.37~\TeV{} in the dielectron mass spectrum with a local significance of 2.5~$\sigma$, but globally the excess is not significant.
The {\textsc{BumpHunter}}~\cite{Choudalakis:2011qn} test, which scans the mass spectrum with varying intervals to find the most significant excess in data, finds $p$-values of \BHpvalee\ and \BHpvalmm\ in the dielectron and dimuon channels, respectively. Figure~\ref{BH_Interesting} shows the dilepton mass distribution in the dielectron and dimuon channels with the observed data overlaid on the combined background prediction, and also the local significance. The interval with the largest upward deviation is indicated by a pair of blue lines.

\begin{figure}[h]
\centering
\includegraphics[width=0.49\linewidth]{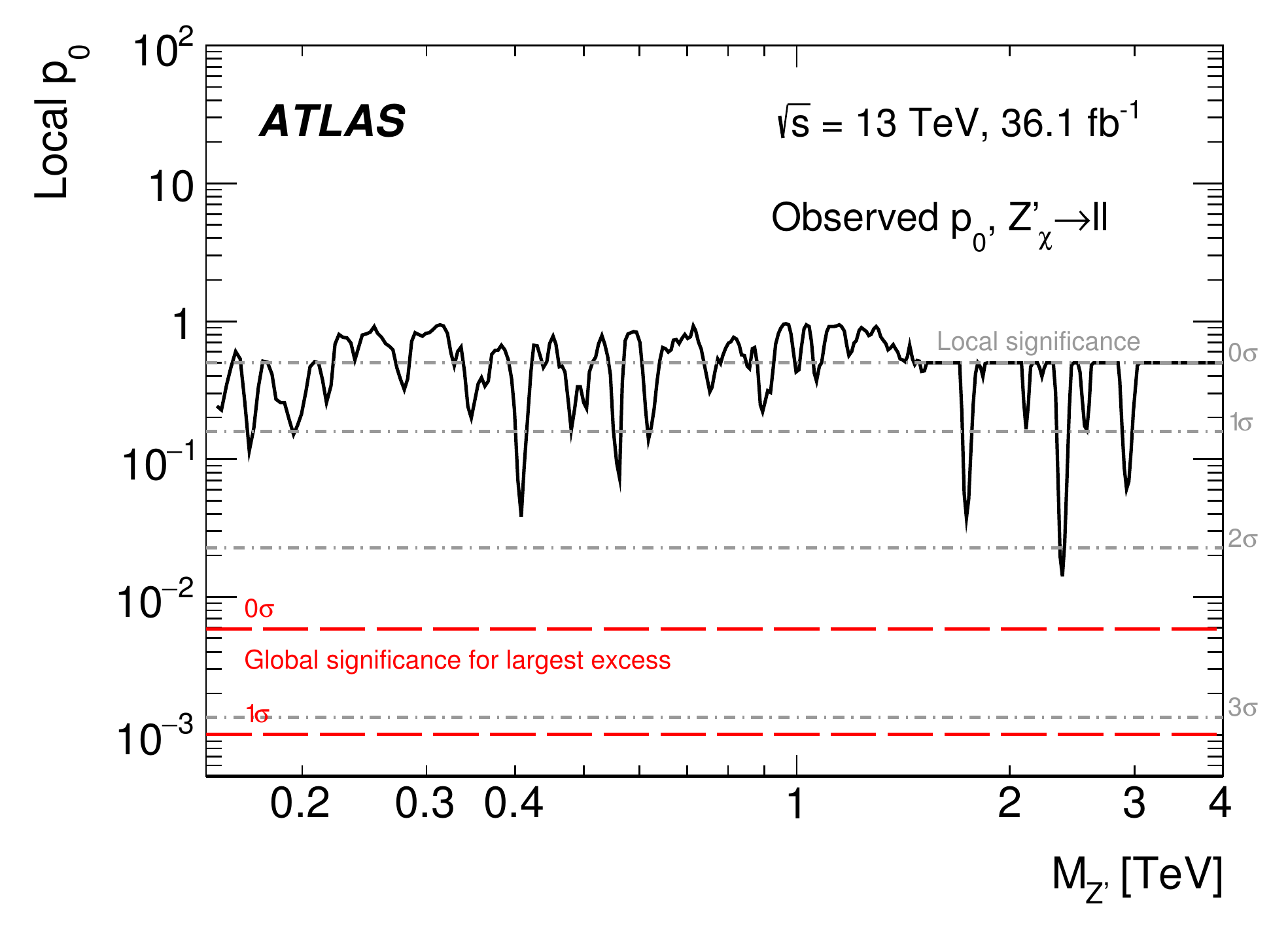}
\caption{The local $p$-value derived assuming \zpchi\ signal shapes with pole masses between 0.15 and 4.0~TeV for the combined dilepton channel. Accompanying local and global significance levels are shown as dashed lines. The {\textit{uncapped}} $p_0$ value is used for pole masses below 1.5~\TeV{}, while the {\textit{capped}} $p_0$ value is used for higher pole masses.
}
\label{fig:observedSignificances}
\end{figure}

\begin{figure}[htb]
\begin{center}
  \subfloat[][]{
    \includegraphics[width=0.49\linewidth]{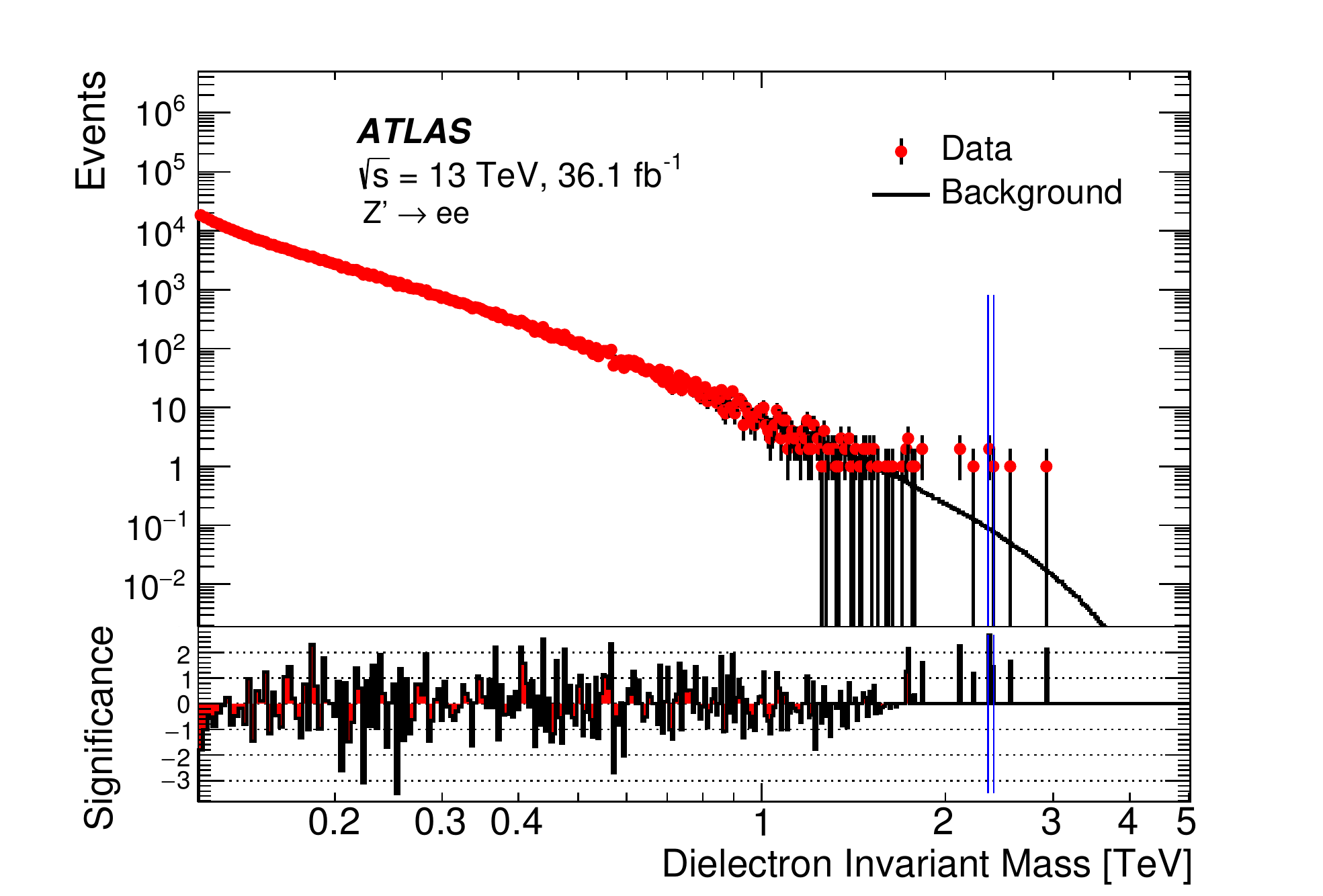}\label{plots_BH_ee}
  }
  \subfloat[][]{
    \includegraphics[width=0.49\linewidth]{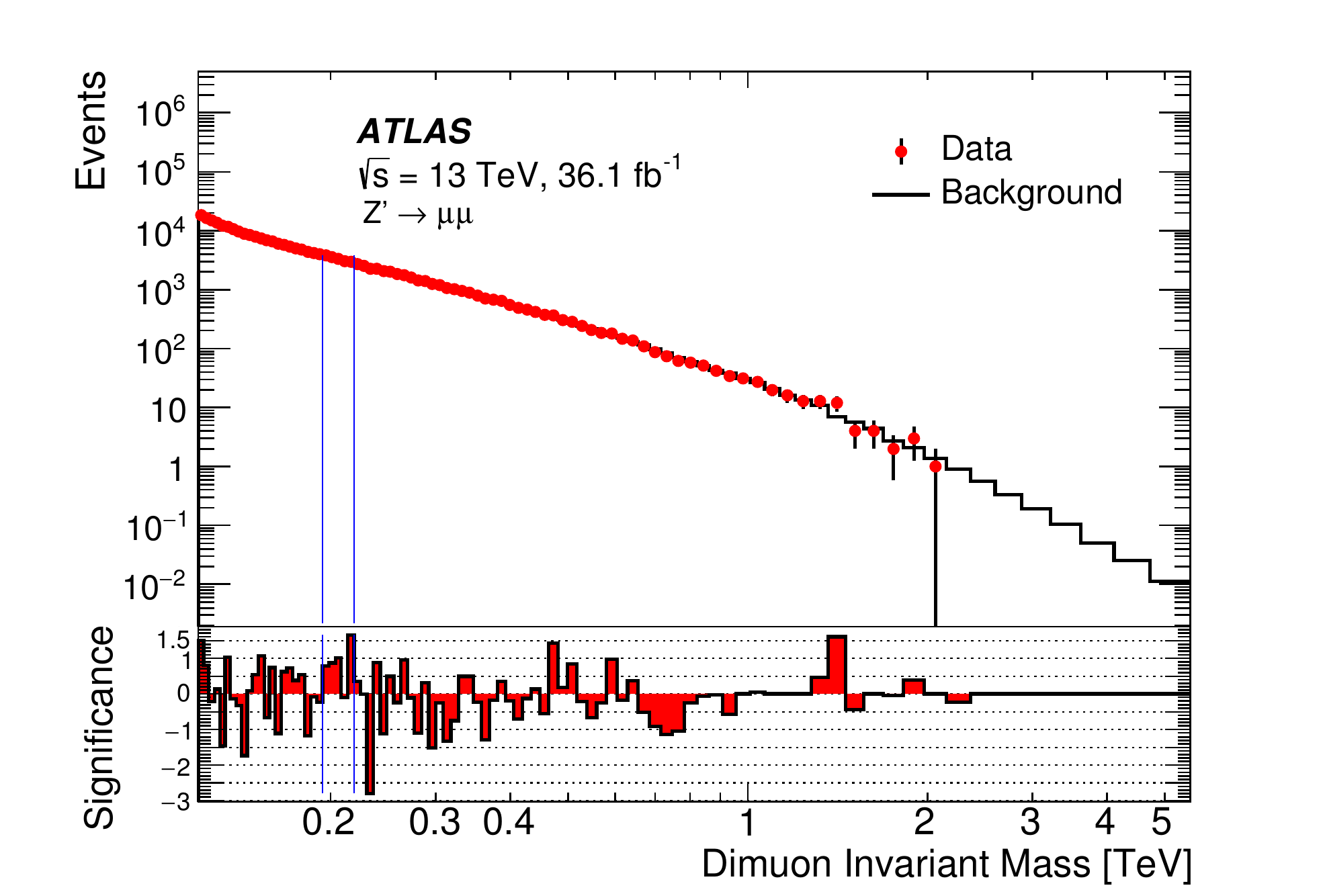}\label{plots_BH_mm}
  }
\end{center}
\vspace{-.4cm}
\caption{Dilepton mass distribution in the \protect\subref{plots_BH_ee} dielectron and \protect\subref{plots_BH_mm} dimuon channel, showing the observed data together with their statistical uncertainty, combined background prediction, and corresponding bin-by-bin significance. The most significant interval is indicated by the vertical blue lines. Exact bin edges and contents are provided in \refT{ee_search_table} and \refT{mm_search_table} in the appendix.}
\label{BH_Interesting}
\end{figure}

\clearpage

\subsection{\zp\ cross-section and mass limits}

Upper limits on the cross-section times branching ratio (\xbr) for \zp\ bosons are presented in Figure~\ref{fig-limits-Zp}. The observed and expected lower limits on the pole mass for various \zp\ scenarios, as described in Section~\ref{theory:e6}, are summarised in \refT{tab-limits-zp}. The \zpchi\ signal is used to extract the limits, which is over-conservative for the other \esix\ models presented, but slightly under-conservative for the \zpssm, although only by 100~\GeV{} in the mass limit at most. The upper limits on \xbr\ for \zp\ bosons start to weaken above a pole mass of $\sim$~3.5~\TeV. The effect is more pronounced in the dimuon channel due to worse mass resolution than in the dielectron channel. The weakening is mainly due to the combined effect of a rapidly falling signal cross-section as the kinematic limit is approached, with an increasing proportion of the signal being produced off-shell in the low-mass tail, and the natural width of the resonance. The selection efficiency also starts to slowly decrease at very high pole masses, but this is a subdominant effect.

\begin{figure}[h]
\begin{center}
\includegraphics[width=0.49\linewidth]{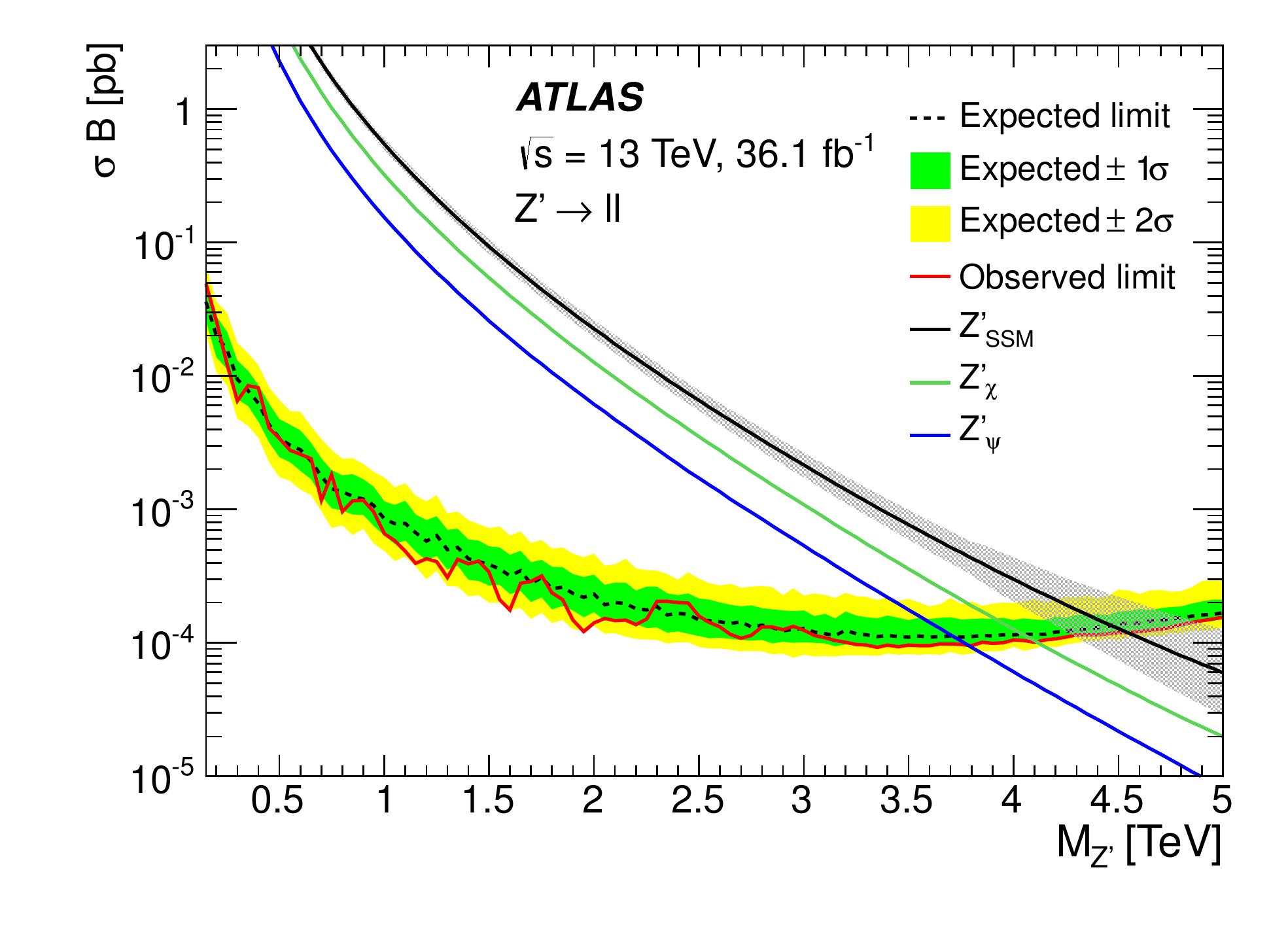}
\end{center}
\vspace{-.4cm}
\caption{Upper 95\% CL limits on the \zp\ production cross-section times branching ratio to two leptons of a single flavour as a function of \zp\ pole mass (M$_{\zp}$). Results are shown for the combined dilepton channel. The signal theoretical \xbr\ are calculated with \pythia~8 using the NNPDF23LO PDF set~\cite{Ball:2012cx}, and corrected to next-to-next-to-leading order in QCD using \vrap~\cite{Anastasiou:2003ds} and the CT14NNLO PDF set~\cite{Dulat:2015mca}. The signals theoretical uncertainties are shown as a band on the \zpssm\ theory line for illustration purposes, but are not included in the \xbr\ limit calculation.}
\label{fig-limits-Zp}
\end{figure}

\begin{table}[h]
  \begin{center}
    \caption{Observed and expected 95\% CL lower mass limits for various \zp\ gauge boson models. The widths are quoted as a percentage of the resonance mass.
    \label{tab-limits-zp}}
    \centering
    \addtolength{\tabcolsep}{+1pt}
    \vspace{0.2cm}
    \begin{tabular}{c|c| r@{$.$}l |cc|cc|cc}
      \hline
      \hline
      \multirow{3}{*}{Model} & \multirow{3}{*}{Width [\%]} & \multicolumn{2}{c|}{\multirow{3}{*}{\te6 [rad]}} & \multicolumn{6}{c}{Lower limits on M$_{\zp}$ [\TeV{}]} \\
      \cline{5-10}
                             &                             & \multicolumn{2}{c|}{}                            & \multicolumn{2}{c|}{$ee$} & \multicolumn{2}{c|}{$\mu\mu$} & \multicolumn{2}{c}{$\ell\ell$} \\
      \cline{5-10}
                             &                             & \multicolumn{2}{c|}{}                            & Obs & Exp &                    Obs & Exp &                   Obs & Exp \\
      \hline
      \zpssm\                & 3.0                         & \multicolumn{2}{c|}{-}                           & \zpeessmobs\ & \zpeessmexp\ & \zpmmssmobs\ & \zpmmssmexp\ & \zpllssmobs\ & \zpllssmexp\ \\
      \zpchi\                & 1.2                         &    0&50~$\pi$                  & \zpeechiobs\ & \zpeechiexp\ & \zpmmchiobs\ & \zpmmchiexp\ & \zpllchiobs\ & \zpllchiexp\ \\
      \zpsq\                 & 1.2                         &    0&63~$\pi$                  & 3.9 & 3.8 & 3.6 & 3.5 & 4.0 & 4.0 \\
      \zpI\                  & 1.1                         &    0&71~$\pi$                  & 3.8 & 3.8 & 3.5 & 3.4 & 4.0 & 3.9 \\
      \zpeta\                & 0.6                         &    0&21~$\pi$                  & 3.7 & 3.7 & 3.4 & 3.3 & 3.9 & 3.8 \\
      \zpN\                  & 0.6                         & $-$0&08~$\pi$                  & 3.6 & 3.6 & 3.4 & 3.3 & 3.8 & 3.8 \\
      \zppsi\                & 0.5                         & \multicolumn{2}{c|}{0~$\pi$}  & \zpeepsiobs\ & \zpeepsiexp\ & \zpmmpsiobs\ & \zpmmpsiexp\ & \zpllpsiobs\ & \zpllpsiexp\ \\
      \hline
      \hline
    \end{tabular}
  \end{center}
\end{table}

\subsection{Limits on Minimal $Z^{\prime}$ models}

Limits are set on the relative coupling strength of the $Z^{\prime}$ boson relative to that of the SM $Z$ boson ($\gamma^{\prime}$) as a function of the $Z^{\prime}_{\text{Min}}$ boson mass, and as a function of the mixing angle $\theta_{\text{Min}}$, as shown in Figure~\ref{fig:minimalModelMassThetaLimits}, and described in Section~\ref{theory:mm}. The two $\theta_{\text{Min}}$ values yielding the minimum and maximum cross-sections are used to define a band of limits in the ($\gamma^{\prime}$, $M_{Z_{\text{Min}}}$) plane. It is possible to put lower mass limits on specific models which are covered by the ($\gamma^{\prime}$, $\theta_{\text{Min}}$) parameterisation as in Table~\ref{tab:SpecificZPrimeMinMassLimits}. The structure observed in the limits as a function of $\theta_{\text{Min}}$, such as the maximum around $\theta_{\text{Min}}$ $=$ 2.2, is due to the changing shape of the resonance at a given pole mass, from narrow to wide.

\begin{figure}[htb]
\begin{center}
  \subfloat[][]{
\includegraphics[width=0.49\linewidth]{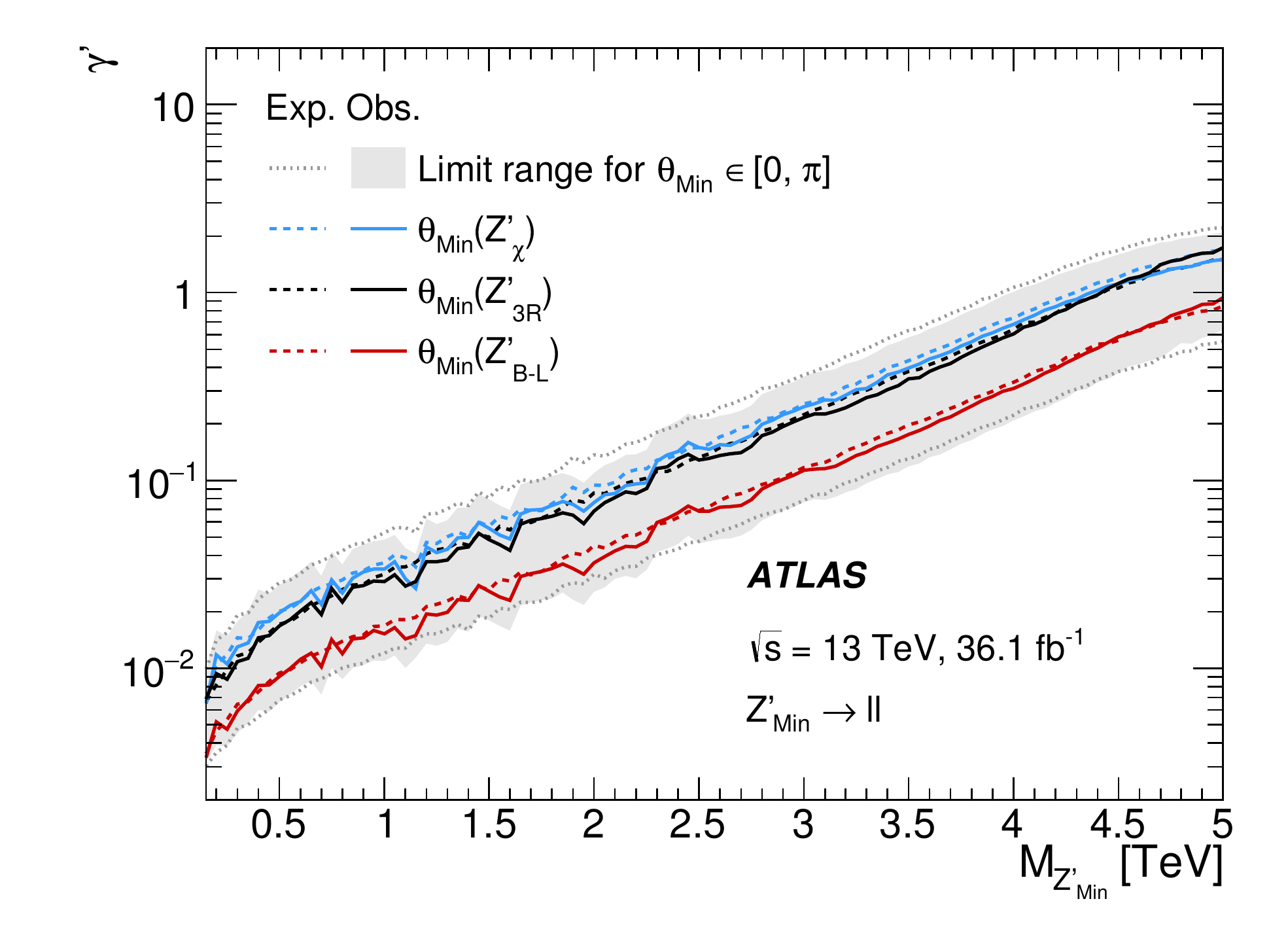}\label{gcoupling_mass_ll}
  }
  \subfloat[][]{
\includegraphics[width=0.49\linewidth]{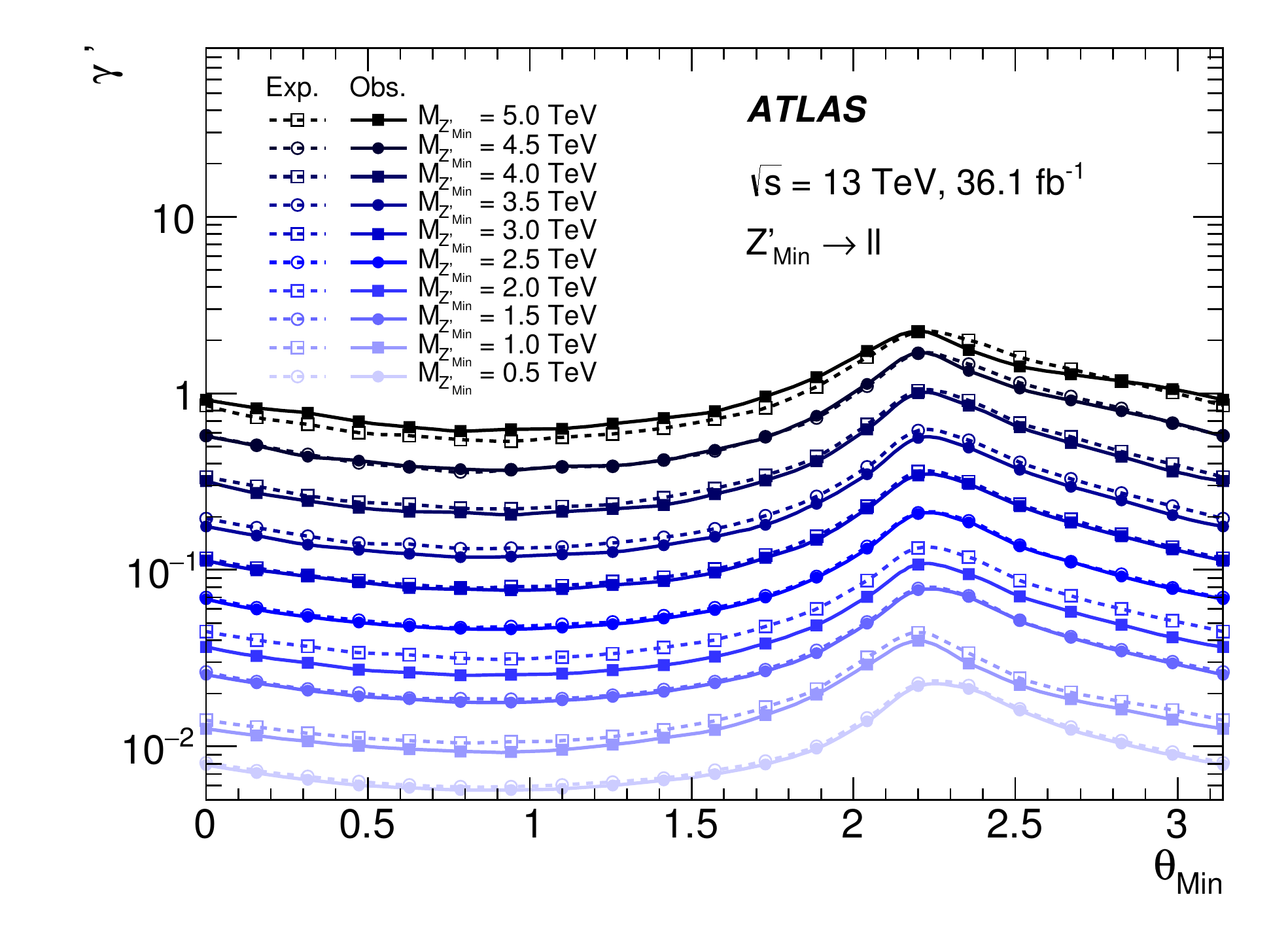}\label{gcoupling_theta_ll}
  }
\end{center}
\vspace{-.4cm}
\caption{
\protect\subref{gcoupling_mass_ll} Expected (dotted and dashed lines) and observed (filled area and lines) limits are set at $95\%$ CL on the relative coupling strength $\gamma^{\prime}$ for the dilepton channel as a function of the $Z^{\prime}_{\text{Min}}$ mass in the Minimal $Z^{\prime}$ model. Limit curves are shown for three representative values of the mixing angle, $\theta_{\text{Min}}$, between the generators of the $\mathrm{(B-L)}$ and the weak hypercharge $Y$ gauge groups. These are: $\tan\theta_{\text{Min}} = 0$, $\tan\theta_{\text{Min}} = -2$ and $\tan\theta_{\text{Min}} = -0.8$, which correspond respectively to the \zpBL, \zpTR\ and $Z^{\prime}_{\chi}$ models at specific values of $\gamma^{\prime}$. The region above each line is excluded. The grey band envelops all observed limit curves, which depend on the choice of $\theta_{\text{Min}} \in [0,\pi]$. The corresponding expected limit curves are within the area delimited by the two dotted lines. \protect\subref{gcoupling_theta_ll} Expected (empty markers and dashed lines) and observed (filled markers and lines) limits at $95\%$ CL on $\gamma^{\prime}$ for the dilepton channel as a function of $\theta_{\text{Min}}$. The limits are set for several representative values of the mass of the $Z^{\prime}$ boson, $M_{Z^{\prime}_{\text{Min}}}$. The region above each line is excluded.}
\label{fig:minimalModelMassThetaLimits}
\end{figure}

\begin{table}[h]
  \begin{center}
    \caption{Observed and expected $95\%$ CL lower mass limits for various $Z^{\prime}_{\text{Min}}$ models.}
    \label{tab:SpecificZPrimeMinMassLimits}
    \centering
    \addtolength{\tabcolsep}{+1pt}
    \vspace{0.2cm}
    \begin{tabular}{c|c|c|cc|cc|cc}
      \hline
      \hline
      \multirow{3}{*}{Model} & \multirow{3}{*}{$\gamma^{\prime}$} & \multirow{3}{*}{$\tan\theta_{\text{Min}}$} & \multicolumn{6}{c}{Lower limits on $M_{Z^{\prime}_{\text{Min}}}$ [\TeV{}] } \\
      \cline{4-9}
      &               &            & \multicolumn{2}{c|}{$ee$} & \multicolumn{2}{c}{$\mu\mu$} & \multicolumn{2}{c}{$\ell\ell$} \\
      \cline{4-9}
      &               &            & Obs & Exp &                    Obs & Exp &                    Obs & Exp  \\
      \hline
      $Z^{\prime}_{\chi}$  & $\sqrt{\frac{41}{24}}\sin\theta_{\text{Min}}$ & $-\frac{4}{5}$ & 3.7 & 3.7 & 3.4 & 3.3 & 3.9 & 3.8 \\
      $Z^{\prime}_{3R}$   & $\sqrt{\frac{5}{8}}\sin\theta_{\text{Min}}$    & $-$2             & 4.0 & 3.9 & 3.6 & 3.6 & 4.1 & 4.1 \\
      $Z^{\prime}_{B - L}$ & $\sqrt{\frac{25}{12}}\sin\theta_{\text{Min}}$ & 0              & 4.0 & 4.0 & 3.6 & 3.6 & 4.2 & 4.1 \\
      \hline
      \hline
    \end{tabular}
    \label{tab:SpecificZPrimeMinMassLimits}
  \end{center}
\end{table}

\clearpage

\subsection{Generic \zp\ limits}

In order to derive more general limits, an approach which compares the data to signals that are more model-independent was developed. This was achieved by applying fiducial cuts to the signal (lepton \pt $>$ 30~\GeV{}, and lepton $|\myeta|$ $<$ 2.5) and a mass window of two times the true signal width (width of the Breit--Wigner) around the pole mass of the signal. This is expected to give limits that are more model independent since any effect on the sensitivity due to the tails of the resonance, foremost the parton luminosity tail and interference effects, are removed. The resulting limits can be seen in Figure~\ref{ModelIndepResultsMW}. For other models to be interpreted with these cross-section limits, the acceptance for a given model in the same fiducial region should be calculated, multiplied by the total cross-section, and the resulting acceptance-corrected cross-section theory curve overlaid, to extract the mass limit for that model. The dilepton invariant mass shape, and angular distributions for the chosen model, should be sufficiently close to a generic \zp\ resonance, such as those presented in this article, so as not to induce additional efficiency differences.

\begin{figure}[htb]
\begin{center}
  \subfloat[][]{
\includegraphics[width=0.49\linewidth]{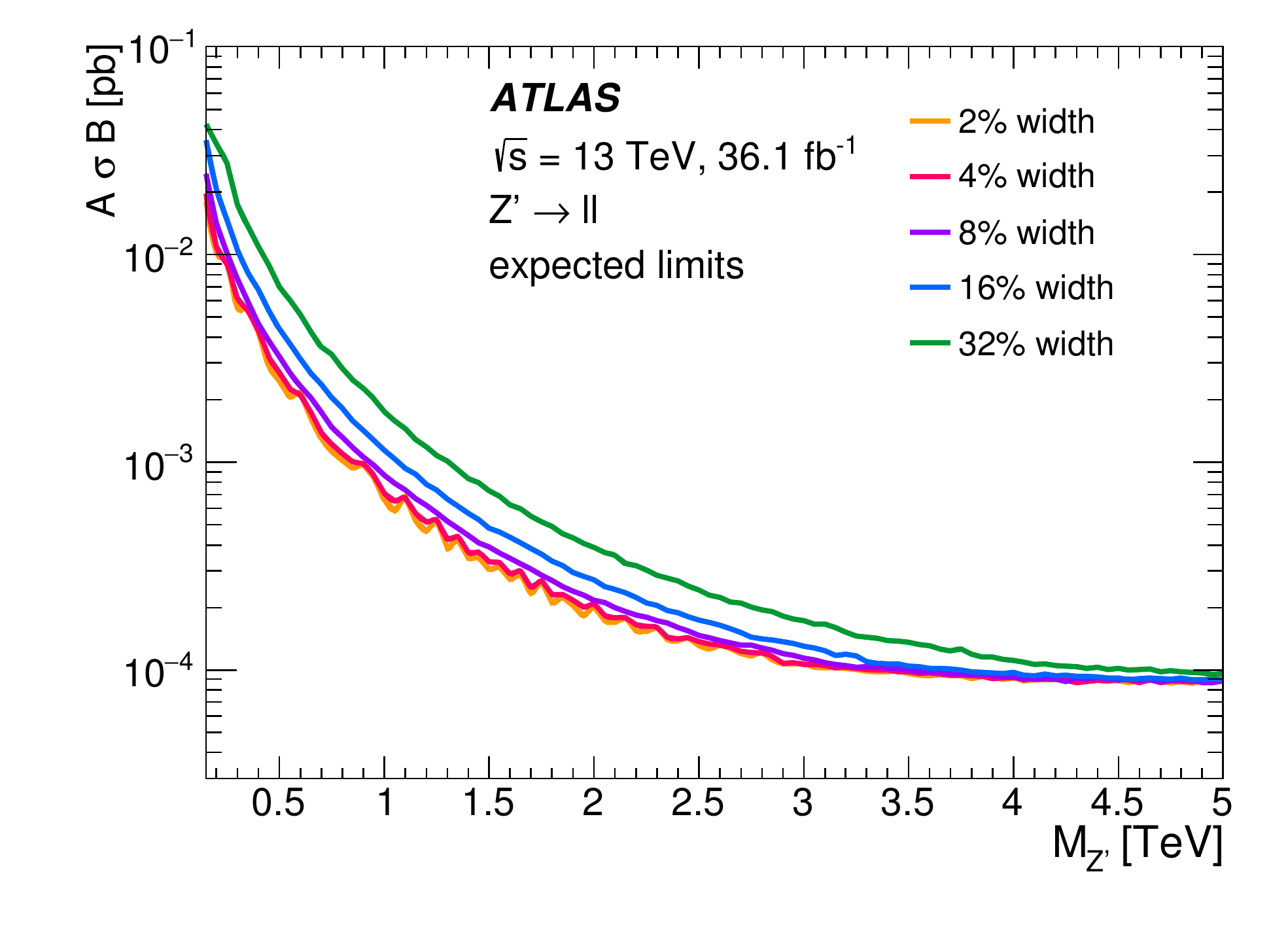}\label{generic_comb_exp}
  }
  \subfloat[][]{
\includegraphics[width=0.49\linewidth]{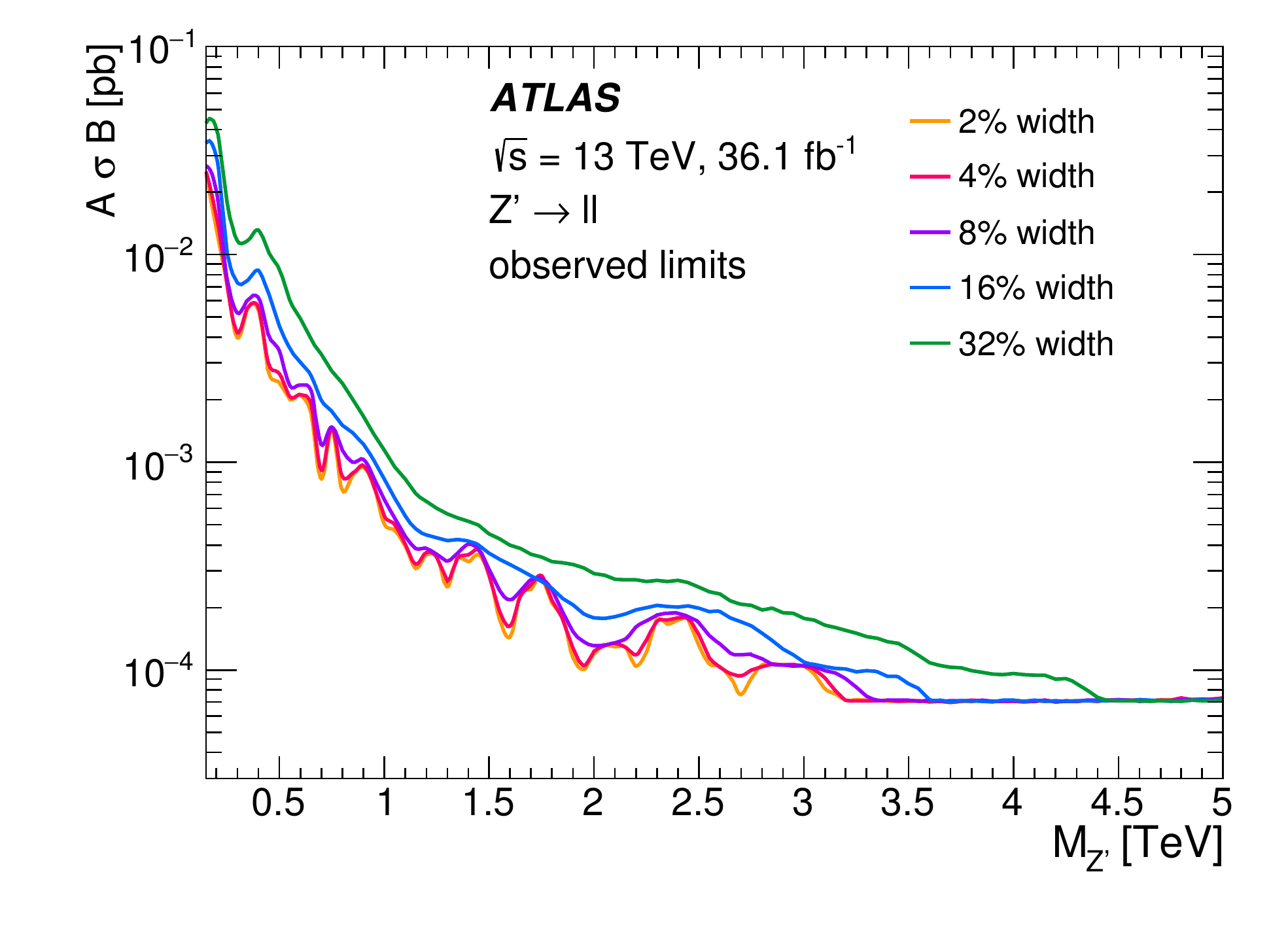}\label{generic_comb_obs}
  }
\end{center}
\vspace{-.4cm}
\caption{Upper 95\% CL limits on the acceptance times \zp\ production cross-section times branching ratio to two leptons of a single flavour as a function of \zp\ pole mass (M$_{\zp}$). \protect\subref{generic_comb_exp} Expected and \protect\subref{generic_comb_obs} observed limits in the combined dilepton channel for different widths with an applied mass window of two times the true width of the signal around the pole mass.}
\label{ModelIndepResultsMW}
\end{figure}

\subsection{Limits on the energy scale of Contact Interactions}

Lower limits are set at 95\% CL on the energy scale $\Lambda$, for the LL, LR, RL, and RR Contact Interaction model, as described in Section~\ref{theory:CI}. Both the constructive and destructive interference scenarios are explored, as well as priors of 1/$\Lambda^2$ and 1/$\Lambda^4$. Limits are presented for the combined dilepton channel in Figure~\ref{CILimitsll} using a 1/$\Lambda^2$ prior. All of the CI exclusion limits are summarised in Table~\ref{tab-limits-ci}.

\begin{figure}[htb]
\begin{center}
\includegraphics[width=0.60\linewidth]{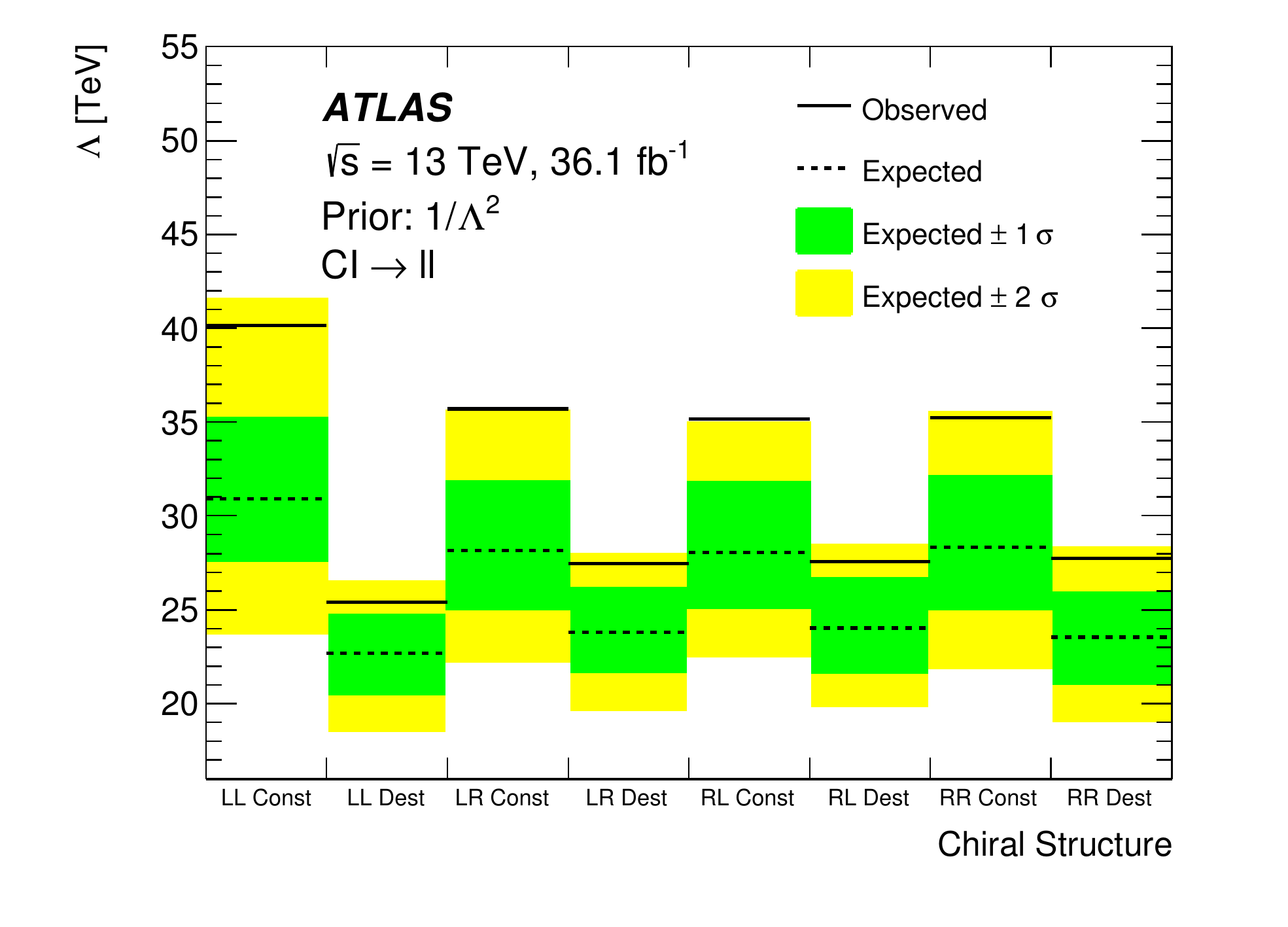}
\end{center}
\vspace{-.4cm}
\caption{Lower limits on the energy scale $\Lambda$ at 95\% CL, for the Contact Interaction model with constructive (const) and destructive (dest) interference, and all considered chiral structures with left-handed (L) and right-handed (R) couplings. Results are shown for the combined dilepton channel.}
\label{CILimitsll}
\end{figure}

\begin{table}[h]
\newcommand\T{\rule{0pt}{2.6ex}}
\newcommand\B{\rule[-1.2ex]{0pt}{0pt}}
\caption{Observed and expected 95\% CL lower limits on $\Lambda$ for the LL, LR, RL, and RR chiral coupling scenarios, for both the constructive (const) and destructive (dest) interference cases using a uniform positive prior in 1/$\Lambda^2$ or 1/$\Lambda^4$. The dielectron, dimuon, and combined dilepton channel limits are shown, rounded to two significant figures.}
\label{tab-limits-ci}
\vspace{0.2cm}
\centering
\begin{tabular}{l|c|c|c|c|c|c|c|c|c}
\hline
\hline
\multirow{3}{*}{Channel} & \multirow{3}{*}{Prior} & \multicolumn{8}{c}{Lower limits on $\Lambda$ [\TeV{}{}]} \T\B \\
\cline{3-10}
  & & \multicolumn{2}{c|}{Left--Left} \T & \multicolumn{2}{c|}{Left--Right} & \multicolumn{2}{c|}{Right--Left} & \multicolumn{2}{c}{Right--Right} \\
\cline{3-10}
& & Const \T & Dest & Const & Dest & Const & Dest & Const & Dest \\
\hline
\hline
Obs: $ee$ \T & \multirow{2}{*}{1/$\Lambda^2$} & 37 & 24 & 33 & 26 & 33 & 26 & 33 & 26 \\
Exp: $ee$ &  & 28 & 22 & 26 & 23 & 26 & 23 & 25 & 23 \\
\hline
Obs: $ee$ \T & \multirow{2}{*}{1/$\Lambda^4$} & 32 & 22 & 29 & 24 & 29 & 24 & 29 & 24 \\
Exp: $ee$ &  & 26 & 20 & 24 & 21 & 24 & 21 & 24 & 21 \\
\hline \hline
Obs: $\mu\mu$ \T & \multirow{2}{*}{1/$\Lambda^2$} & 30 & 20 & 28 & 22 & 28 & 22 & 28 & 20 \\
Exp: $\mu\mu$ &  & 26 & 20 & 24 & 21 & 24 & 21 & 24 & 20 \\
\hline
Obs: $\mu\mu$ \T & \multirow{2}{*}{1/$\Lambda^4$} & 27 & 19 & 25 & 21 & 25 & 21 & 25 & 19 \\
Exp: $\mu\mu$ &  & 24 & 18 & 23 & 20 & 22 & 20 & 22 & 18 \\
\hline \hline
Obs: $\ell\ell$ \T & \multirow{2}{*}{1/$\Lambda^2$}  & 40 & 25 & 36 & 28 & 35 & 28 & 35 & 28 \\
Exp: $\ell\ell$ &  & 31 & 23 & 28 & 24 & 28 & 24 & 28 & 24 \\
\hline
Obs: $\ell\ell$ \T & \multirow{2}{*}{1/$\Lambda^4$} & 35 & 24 & 32 & 25 & 32 & 25 & 31 & 25 \\
Exp: $\ell\ell$ &  & 28 & 21 & 26 & 22 & 26 & 23 & 26 & 22 \\
\hline
\hline
\end{tabular}
\end{table}

\cleardoublepage


\FloatBarrier

\section{Conclusion}
\label{sec:conclusion}
The ATLAS detector at the LHC has been used to search for both resonant and non-resonant new phenomena in the dilepton invariant mass spectrum above the $Z$ boson's pole. The search is conducted with \lumill\ of $pp$ collision data at $\sqrt{s}$ = 13~\TeV{}, recorded during 2015 and 2016.
The highest invariant mass event is found at \HighestMee~\TeV{} in the dielectron channel, and \HighestMmumu~\TeV{} in the dimuon channel.
The observed dilepton invariant mass spectrum is consistent with the Standard Model prediction, within systematic and statistical uncertainties.
Among a choice of different models, the data are interpreted in terms of resonant spin-1 \zp\ gauge boson production and non-resonant $qq \ell\ell$ contact interactions.
For the resonant interpretation, upper limits are set on the cross-section times branching ratio for a spin-1 \zp\ gauge boson.
The resulting 95\% CL lower mass limits are \zpllssmobs~\TeV{} for the \zpssm, \zpllchiobs~\TeV{} for the \zpchi, and \zpllpsiobs~\TeV{} for the \zppsi.
Other \esix\ \zp\ models are also constrained in the range between those quoted for the \zpchi\ and \zppsi.
This result is more stringent than the previous ATLAS result at $\sqrt{s}$ = 13~\TeV{} obtained with 2015 data, by up to 700~\GeV{}.
Lower mass limits are also set on the Minimal \zp\ model, up to 4.1~\TeV{} for the \zpTR, and 4.2~\TeV{} for the \zpBL.
Generic \zp\ cross-section limits are also provided for a range of true signal widths.
The lower limits on the energy scale $\Lambda$ for various $qq \ell\ell$ contact interaction models range between \CIllobsLow~\TeV{} and \CIllobsHigh~\TeV{}, which are more stringent than the previous ATLAS result obtained at $\sqrt{s}$ = 13~\TeV{}, by up to 4~\TeV{}.


\clearpage
\appendix
\part*{Appendix}
This appendix provides the exact bin edges and contents of the dilepton invariant mass plots presented in Figures~\ref{plots_inv_mass_ee}, \ref{plots_inv_mass_mm}, \ref{plots_BH_ee}, and \ref{plots_BH_mm}. These correspond to Tables~\ref{ee_log_table}, \ref{mm_log_table}, \ref{ee_search_table}, and \ref{mm_search_table}, respectively. Even more detailed information can be found in the Durham HEP database~\cite{HepData}.

{\footnotesize
\begin{longtable}{llll}
\caption{Expected and observed event yields in the dielectron channel, directly corresponding to the non-linear binning presented in Figure~\ref{plots_inv_mass_ee}. The expected yield is given up to at most 4 digit precision.} \label{ee_log_table} \\
\toprule
\toprule
Lower edge [\GeV] & Upper edge [\GeV] & Data [N] & Total Background [N] \\
\midrule
80 & 85.549 & 1176847 & 1112000\\
85.549 & 91.482 & 6608874 & 6322000\\
91.482 & 97.828 & 3928394 & 3756000\\
97.828 & 104.61 & 432217 & 414400\\
104.61 & 111.87 & 162962 & 156100\\
111.87 & 119.63 & 93773 & 90620\\
119.63 & 127.93 & 63446 & 62270\\
127.93 & 136.8 & 47190 & 46740\\
136.8 & 146.29 & 36539 & 36090\\
146.29 & 156.43 & 29267 & 28990\\
156.43 & 167.28 & 23874 & 23740\\
167.28 & 178.89 & 19689 & 19550\\
178.89 & 191.29 & 16548 & 16400\\
191.29 & 204.56 & 13671 & 13590\\
204.56 & 218.75 & 11337 & 11460\\
218.75 & 233.92 & 9358 & 9499\\
233.92 & 250.15 & 7877 & 7868\\
250.15 & 267.5 & 6434 & 6570\\
267.5 & 286.05 & 5500 & 5427\\
286.05 & 305.89 & 4445 & 4477\\
305.89 & 327.11 & 3648 & 3667\\
327.11 & 349.79 & 2981 & 2995\\
349.79 & 374.06 & 2431 & 2403\\
374.06 & 400 & 1964 & 1957\\
400 & 427.74 & 1606 & 1565\\
427.74 & 457.41 & 1231 & 1265\\
457.41 & 489.14 & 1013 & 1008\\
489.14 & 523.06 & 776 & 805.6\\
523.06 & 559.34 & 622 & 628.7\\
559.34 & 598.14 & 464 & 492.3\\
598.14 & 639.63 & 403 & 392.6\\
639.63 & 683.99 & 300 & 304.4\\
683.99 & 731.43 & 219 & 234.3\\
731.43 & 782.16 & 202 & 183.2\\
782.16 & 836.41 & 133 & 140.2\\
836.41 & 894.43 & 107 & 107.1\\
894.43 & 956.46 & 82 & 85.13\\
956.46 & 1022.8 & 57 & 63.86\\
1022.8 & 1093.7 & 43 & 47.9\\
1093.7 & 1169.6 & 27 & 38.09\\
1169.6 & 1250.7 & 24 & 28.7\\
1250.7 & 1337.5 & 12 & 20.28\\
1337.5 & 1430.2 & 13 & 14.96\\
1430.2 & 1529.4 & 11 & 11.16\\
1529.4 & 1635.5 & 3 & 8.262\\
1635.5 & 1749 & 7 & 6.003\\
1749 & 1870.3 & 4 & 4.085\\
1870.3 & 2000 & 0 & 2.875\\
2000 & 2138.7 & 2 & 2.05\\
2138.7 & 2287.1 & 1 & 1.431\\
2287.1 & 2445.7 & 3 & 0.977\\
2445.7 & 2615.3 & 1 & 0.655\\
2615.3 & 2796.7 & 0 & 0.443\\
2796.7 & 2990.7 & 1 & 0.284\\
2990.7 & 3198.1 & 0 & 0.183\\
3198.1 & 3420 & 0 & 0.114\\
3420 & 3657.2 & 0 & 0.068\\
3657.2 & 3910.8 & 0 & 0.041\\
3910.8 & 4182.1 & 0 & 0.023\\
4182.1 & 4472.1 & 0 & 0.013\\
4472.1 & 4782.3 & 0 & 0.007\\
4782.3 & 5114 & 0 & 0.004\\
5114 & 5468.7 & 0 & 0.002\\
5468.7 & 5848 & 0 & 0.001\\
5848 & 6253.7 & 0 & 0\\
\bottomrule
\bottomrule
\end{longtable}

\begin{longtable}{llll}
\caption{Expected and observed event yields in the dimuon channel, directly corresponding to the non-linear binning presented in Figure~\ref{plots_inv_mass_mm}. The expected yield is given up to at most 4 digit precision.} \label{mm_log_table} \\
\toprule
\toprule
Lower edge [\GeV] & Upper edge [\GeV] & Data [N] & Total Background [N] \\
\midrule
80 & 85.549 & 826504 & 786600\\
85.549 & 91.482 & 5730639 & 5465000\\
91.482 & 97.828 & 4062661 & 3848000\\
97.828 & 104.61 & 430822 & 405500\\
104.61 & 111.87 & 149927 & 141800\\
111.87 & 119.63 & 82971 & 79230\\
119.63 & 127.93 & 54641 & 52110\\
127.93 & 136.8 & 39501 & 37890\\
136.8 & 146.29 & 29742 & 28940\\
146.29 & 156.43 & 23871 & 23220\\
156.43 & 167.28 & 18942 & 18490\\
167.28 & 178.89 & 15482 & 15140\\
178.89 & 191.29 & 12495 & 12250\\
191.29 & 204.56 & 10462 & 10230\\
204.56 & 218.75 & 8583 & 8261\\
218.75 & 233.92 & 6868 & 6885\\
233.92 & 250.15 & 5649 & 5517\\
250.15 & 267.5 & 4723 & 4607\\
267.5 & 286.05 & 3762 & 3753\\
286.05 & 305.89 & 3064 & 3106\\
305.89 & 327.11 & 2471 & 2566\\
327.11 & 349.79 & 2031 & 1992\\
349.79 & 374.06 & 1595 & 1628\\
374.06 & 400 & 1333 & 1321\\
400 & 427.74 & 1018 & 1022\\
427.74 & 457.41 & 819 & 828.1\\
457.41 & 489.14 & 675 & 651.9\\
489.14 & 523.06 & 508 & 513.6\\
523.06 & 559.34 & 397 & 410.7\\
559.34 & 598.14 & 306 & 306\\
598.14 & 639.63 & 252 & 245.7\\
639.63 & 683.99 & 188 & 191.2\\
683.99 & 731.43 & 129 & 142.2\\
731.43 & 782.16 & 97 & 108.5\\
782.16 & 836.41 & 78 & 82.36\\
836.41 & 894.43 & 57 & 63.72\\
894.43 & 956.46 & 51 & 51.88\\
956.46 & 1022.8 & 39 & 39.04\\
1022.8 & 1093.7 & 29 & 29.74\\
1093.7 & 1169.6 & 18 & 21.83\\
1169.6 & 1250.7 & 18 & 16.12\\
1250.7 & 1337.5 & 14 & 12.7\\
1337.5 & 1430.2 & 12 & 8.053\\
1430.2 & 1529.4 & 5 & 5.803\\
1529.4 & 1635.5 & 4 & 4.667\\
1635.5 & 1749 & 1 & 3.241\\
1749 & 1870.3 & 4 & 2.135\\
1870.3 & 2000 & 2 & 1.663\\
2000 & 2138.7 & 0 & 1.102\\
2138.7 & 2287.1 & 0 & 0.763\\
2287.1 & 2445.7 & 0 & 0.538\\
2445.7 & 2615.3 & 0 & 0.375\\
2615.3 & 2796.7 & 0 & 0.250\\
2796.7 & 2990.7 & 0 & 0.165\\
2990.7 & 3198.1 & 0 & 0.112\\
3198.1 & 3420 & 0 & 0.078\\
3420 & 3657.2 & 0 & 0.049\\
3657.2 & 3910.8 & 0 & 0.031\\
3910.8 & 4182.1 & 0 & 0.022\\
4182.1 & 4472.1 & 0 & 0.013\\
4472.1 & 4782.3 & 0 & 0.010\\
4782.3 & 5114 & 0 & 0.006\\
5114 & 5468.7 & 0 & 0.005\\
5468.7 & 5848 & 0 & 0.002\\
5848 & 6253.7 & 0 & 0.002\\
\bottomrule
\bottomrule
\end{longtable}

\begin{longtable}{llll}
\caption{Expected and observed event yields in the dielectron channel, directly corresponding to the non-linear binning presented in Figure~\ref{plots_BH_ee}. The expected yield is given up to at most 4 digit precision.} \label{ee_search_table} \\
\toprule
\toprule
Lower edge [\TeV] & Upper edge [\TeV] & Data [N] & Total Background [N] \\
\midrule
0.11962 & 0.12171 & 18432 & 18660\\
0.12171 & 0.12381 & 16720 & 16840\\
0.12381 & 0.12592 & 15291 & 15340\\
0.12592 & 0.12806 & 13924 & 14020\\
0.12806 & 0.13022 & 12931 & 12960\\
0.13022 & 0.13239 & 11976 & 11970\\
0.13239 & 0.13459 & 11154 & 11120\\
0.13459 & 0.1368 & 10273 & 10350\\
0.1368 & 0.13903 & 9637 & 9677\\
0.13903 & 0.14128 & 9017 & 9029\\
0.14128 & 0.14355 & 8392 & 8460\\
0.14355 & 0.14584 & 8043 & 7949\\
0.14584 & 0.14815 & 7387 & 7507\\
0.14815 & 0.15048 & 7122 & 7073\\
0.15048 & 0.15283 & 6695 & 6680\\
0.15283 & 0.1552 & 6382 & 6278\\
0.1552 & 0.15758 & 5933 & 6018\\
0.15758 & 0.15999 & 5737 & 5697\\
0.15999 & 0.16242 & 5352 & 5380\\
0.16242 & 0.16487 & 5097 & 5158\\
0.16487 & 0.16733 & 4972 & 4893\\
0.16733 & 0.16982 & 4726 & 4622\\
0.16982 & 0.17233 & 4455 & 4437\\
0.17233 & 0.17486 & 4165 & 4198\\
0.17486 & 0.17741 & 3979 & 4058\\
0.17741 & 0.17998 & 3920 & 3868\\
0.17998 & 0.18257 & 3629 & 3687\\
0.18257 & 0.18518 & 3671 & 3531\\
0.18518 & 0.18782 & 3392 & 3349\\
0.18782 & 0.19047 & 3154 & 3207\\
0.19047 & 0.19315 & 3182 & 3081\\
0.19315 & 0.19584 & 2964 & 2957\\
0.19584 & 0.19856 & 2826 & 2825\\
0.19856 & 0.2013 & 2687 & 2712\\
0.2013 & 0.20406 & 2637 & 2593\\
0.20406 & 0.20685 & 2380 & 2510\\
0.20685 & 0.20965 & 2436 & 2396\\
0.20965 & 0.21248 & 2223 & 2295\\
0.21248 & 0.21533 & 2166 & 2202\\
0.21533 & 0.2182 & 2152 & 2089\\
0.2182 & 0.2211 & 2030 & 2003\\
0.2211 & 0.22402 & 1803 & 1938\\
0.22402 & 0.22696 & 1869 & 1831\\
0.22696 & 0.22992 & 1736 & 1774\\
0.22992 & 0.23291 & 1788 & 1726\\
0.23291 & 0.23591 & 1605 & 1646\\
0.23591 & 0.23895 & 1651 & 1576\\
0.23895 & 0.242 & 1530 & 1515\\
0.242 & 0.24508 & 1386 & 1453\\
0.24508 & 0.24818 & 1398 & 1398\\
0.24818 & 0.25131 & 1354 & 1328\\
0.25131 & 0.25446 & 1171 & 1297\\
0.25446 & 0.25763 & 1293 & 1246\\
0.25763 & 0.26083 & 1139 & 1191\\
0.26083 & 0.26405 & 1187 & 1155\\
0.26405 & 0.2673 & 1075 & 1117\\
0.2673 & 0.27057 & 1051 & 1073\\
0.27057 & 0.27387 & 1045 & 1022\\
0.27387 & 0.27719 & 1011 & 988.8\\
0.27719 & 0.28053 & 953 & 943.4\\
0.28053 & 0.2839 & 966 & 908.3\\
0.2839 & 0.2873 & 834 & 878.5\\
0.2873 & 0.29072 & 820 & 840.8\\
0.29072 & 0.29416 & 813 & 809.1\\
0.29416 & 0.29764 & 799 & 782.9\\
0.29764 & 0.30113 & 723 & 742.1\\
0.30113 & 0.30466 & 746 & 722.8\\
0.30466 & 0.30821 & 692 & 696.2\\
0.30821 & 0.31178 & 655 & 668.4\\
0.31178 & 0.31539 & 639 & 656.2\\
0.31539 & 0.31901 & 607 & 615.3\\
0.31901 & 0.32267 & 594 & 597.5\\
0.32267 & 0.32635 & 586 & 568.5\\
0.32635 & 0.33006 & 560 & 550.5\\
0.33006 & 0.3338 & 524 & 520.3\\
0.3338 & 0.33756 & 475 & 512.2\\
0.33756 & 0.34135 & 500 & 499\\
0.34135 & 0.34517 & 490 & 466.4\\
0.34517 & 0.34902 & 465 & 443.6\\
0.34902 & 0.35289 & 443 & 436.7\\
0.35289 & 0.3568 & 417 & 420.2\\
0.3568 & 0.36073 & 413 & 393.5\\
0.36073 & 0.36468 & 377 & 380.7\\
0.36468 & 0.36867 & 410 & 366.5\\
0.36867 & 0.37269 & 338 & 365.1\\
0.37269 & 0.37673 & 373 & 358.6\\
0.37673 & 0.38081 & 330 & 338.5\\
0.38081 & 0.38491 & 304 & 311.7\\
0.38491 & 0.38905 & 312 & 306.7\\
0.38905 & 0.39321 & 299 & 308.5\\
0.39321 & 0.3974 & 290 & 281.3\\
0.3974 & 0.40162 & 268 & 272.1\\
0.40162 & 0.40588 & 298 & 261.1\\
0.40588 & 0.41016 & 281 & 255.1\\
0.41016 & 0.41447 & 253 & 240\\
0.41447 & 0.41882 & 235 & 237.1\\
0.41882 & 0.42319 & 242 & 228.2\\
0.42319 & 0.4276 & 192 & 216.4\\
0.4276 & 0.43203 & 215 & 210.4\\
0.43203 & 0.4365 & 192 & 204.7\\
0.4365 & 0.441 & 231 & 194.4\\
0.441 & 0.44553 & 171 & 186\\
0.44553 & 0.4501 & 184 & 181\\
0.4501 & 0.45469 & 153 & 172.6\\
0.45469 & 0.45932 & 153 & 168.9\\
0.45932 & 0.46398 & 178 & 156.2\\
0.46398 & 0.46867 & 150 & 157.4\\
0.46867 & 0.4734 & 139 & 158.8\\
0.4734 & 0.47816 & 171 & 144.5\\
0.47816 & 0.48295 & 138 & 143.3\\
0.48295 & 0.48778 & 140 & 131.8\\
0.48778 & 0.49264 & 121 & 127.9\\
0.49264 & 0.49753 & 117 & 127.2\\
0.49753 & 0.50246 & 127 & 119.3\\
0.50246 & 0.50742 & 127 & 114.7\\
0.50742 & 0.51242 & 100 & 113.8\\
0.51242 & 0.51745 & 101 & 105.1\\
0.51745 & 0.52252 & 113 & 105.4\\
0.52252 & 0.52762 & 94 & 100.9\\
0.52762 & 0.53276 & 82 & 95.19\\
0.53276 & 0.53793 & 101 & 90.98\\
0.53793 & 0.54314 & 74 & 87.16\\
0.54314 & 0.54839 & 89 & 85.14\\
0.54839 & 0.55367 & 92 & 81.42\\
0.55367 & 0.55898 & 91 & 81.14\\
0.55898 & 0.56434 & 84 & 78.37\\
0.56434 & 0.56973 & 96 & 74.86\\
0.56973 & 0.57516 & 51 & 73.66\\
0.57516 & 0.58062 & 60 & 65.09\\
0.58062 & 0.58613 & 59 & 66.37\\
0.58613 & 0.59167 & 64 & 63.71\\
0.59167 & 0.59725 & 47 & 63.18\\
0.59725 & 0.60286 & 62 & 62.67\\
0.60286 & 0.60852 & 64 & 59.51\\
0.60852 & 0.61422 & 54 & 51.24\\
0.61422 & 0.61995 & 62 & 52.4\\
0.61995 & 0.62572 & 49 & 49.37\\
0.62572 & 0.63153 & 57 & 50.37\\
0.63153 & 0.63739 & 48 & 48.26\\
0.63739 & 0.64328 & 43 & 48.46\\
0.64328 & 0.64921 & 41 & 45.4\\
0.64921 & 0.65518 & 45 & 42.38\\
0.65518 & 0.6612 & 42 & 40.13\\
0.6612 & 0.66725 & 40 & 40.51\\
0.66725 & 0.67335 & 36 & 36.82\\
0.67335 & 0.67949 & 33 & 38.69\\
0.67949 & 0.68567 & 44 & 34.63\\
0.68567 & 0.69189 & 27 & 34.25\\
0.69189 & 0.69815 & 36 & 35.36\\
0.69815 & 0.70446 & 40 & 31.61\\
0.70446 & 0.71081 & 24 & 30.31\\
0.71081 & 0.7172 & 21 & 28.97\\
0.7172 & 0.72363 & 29 & 28.06\\
0.72363 & 0.73011 & 29 & 26.9\\
0.73011 & 0.73663 & 35 & 26.39\\
0.73663 & 0.7432 & 20 & 25.87\\
0.7432 & 0.74981 & 31 & 24.46\\
0.74981 & 0.75647 & 27 & 24.33\\
0.75647 & 0.76317 & 26 & 23.24\\
0.76317 & 0.76991 & 25 & 22.35\\
0.76991 & 0.77671 & 19 & 21.38\\
0.77671 & 0.78354 & 27 & 20.01\\
0.78354 & 0.79043 & 21 & 21\\
0.79043 & 0.79736 & 15 & 18.29\\
0.79736 & 0.80433 & 17 & 18.3\\
0.80433 & 0.81136 & 22 & 17.46\\
0.81136 & 0.81843 & 13 & 17.52\\
0.81843 & 0.82555 & 17 & 19.08\\
0.82555 & 0.83271 & 17 & 16.08\\
0.83271 & 0.83993 & 18 & 15.5\\
0.83993 & 0.84719 & 12 & 14.39\\
0.84719 & 0.8545 & 14 & 14.37\\
0.8545 & 0.86186 & 20 & 12.71\\
0.86186 & 0.86927 & 9 & 13.61\\
0.86927 & 0.87673 & 8 & 12.34\\
0.87673 & 0.88424 & 17 & 12.77\\
0.88424 & 0.8918 & 10 & 12.03\\
0.8918 & 0.89941 & 19 & 11.72\\
0.89941 & 0.90708 & 8 & 11\\
0.90708 & 0.91479 & 12 & 10.91\\
0.91479 & 0.92255 & 14 & 10.52\\
0.92255 & 0.93037 & 13 & 10.63\\
0.93037 & 0.93824 & 5 & 9.507\\
0.93824 & 0.94616 & 10 & 9.246\\
0.94616 & 0.95413 & 8 & 9.899\\
0.95413 & 0.96216 & 6 & 9.358\\
0.96216 & 0.97024 & 7 & 9.246\\
0.97024 & 0.97838 & 5 & 8.622\\
0.97838 & 0.98657 & 7 & 7.47\\
0.98657 & 0.99481 & 9 & 8.008\\
0.99481 & 1.0031 & 8 & 6.946\\
1.0031 & 1.0115 & 10 & 6.434\\
1.0115 & 1.0199 & 5 & 6.917\\
1.0199 & 1.0283 & 4 & 6.297\\
1.0283 & 1.0369 & 3 & 6.431\\
1.0369 & 1.0454 & 4 & 5.779\\
1.0454 & 1.0541 & 5 & 6.033\\
1.0541 & 1.0628 & 9 & 5.665\\
1.0628 & 1.0715 & 7 & 5.245\\
1.0715 & 1.0803 & 3 & 5.998\\
1.0803 & 1.0892 & 6 & 4.778\\
1.0892 & 1.0981 & 6 & 4.841\\
1.0981 & 1.1071 & 2 & 4.69\\
1.1071 & 1.1162 & 3 & 4.427\\
1.1162 & 1.1253 & 4 & 4.615\\
1.1253 & 1.1344 & 3 & 4.067\\
1.1344 & 1.1437 & 3 & 4.272\\
1.1437 & 1.1529 & 3 & 4.283\\
1.1529 & 1.1623 & 2 & 4.086\\
1.1623 & 1.1717 & 4 & 4.619\\
1.1717 & 1.1812 & 3 & 3.411\\
1.1812 & 1.1907 & 6 & 3.524\\
1.1907 & 1.2003 & 2 & 3.687\\
1.2003 & 1.21 & 2 & 3.449\\
1.21 & 1.2197 & 5 & 3.116\\
1.2197 & 1.2295 & 0 & 3.312\\
1.2295 & 1.2393 & 3 & 3.577\\
1.2393 & 1.2493 & 2 & 2.714\\
1.2493 & 1.2593 & 1 & 2.86\\
1.2593 & 1.2693 & 4 & 2.6\\
1.2693 & 1.2794 & 1 & 2.416\\
1.2794 & 1.2896 & 2 & 2.339\\
1.2896 & 1.2999 & 0 & 2.305\\
1.2999 & 1.3102 & 2 & 2.494\\
1.3102 & 1.3206 & 2 & 2.128\\
1.3206 & 1.331 & 1 & 1.947\\
1.331 & 1.3416 & 1 & 1.946\\
1.3416 & 1.3522 & 3 & 1.999\\
1.3522 & 1.3628 & 0 & 1.673\\
1.3628 & 1.3736 & 2 & 1.917\\
1.3736 & 1.3844 & 0 & 1.753\\
1.3844 & 1.3953 & 3 & 1.592\\
1.3953 & 1.4062 & 1 & 1.503\\
1.4062 & 1.4172 & 1 & 1.43\\
1.4172 & 1.4283 & 2 & 1.445\\
1.4283 & 1.4395 & 0 & 1.361\\
1.4395 & 1.4507 & 1 & 1.415\\
1.4507 & 1.462 & 1 & 1.253\\
1.462 & 1.4734 & 2 & 1.2\\
1.4734 & 1.4849 & 2 & 1.389\\
1.4849 & 1.4964 & 0 & 1.177\\
1.4964 & 1.5081 & 2 & 1.17\\
1.5081 & 1.5198 & 1 & 1.142\\
1.5198 & 1.5315 & 2 & 0.959\\
1.5315 & 1.5434 & 0 & 1.074\\
1.5434 & 1.5553 & 1 & 0.878\\
1.5553 & 1.5673 & 0 & 1.037\\
1.5673 & 1.5794 & 0 & 0.843\\
1.5794 & 1.5915 & 0 & 0.787\\
1.5915 & 1.6038 & 1 & 0.922\\
1.6038 & 1.6161 & 1 & 0.767\\
1.6161 & 1.6285 & 0 & 0.873\\
1.6285 & 1.641 & 1 & 0.711\\
1.641 & 1.6536 & 0 & 0.683\\
1.6536 & 1.6662 & 0 & 0.808\\
1.6662 & 1.6789 & 0 & 0.682\\
1.6789 & 1.6918 & 0 & 0.627\\
1.6918 & 1.7047 & 1 & 0.702\\
1.7047 & 1.7176 & 0 & 0.562\\
1.7176 & 1.7307 & 2 & 0.569\\
1.7307 & 1.7439 & 3 & 0.530\\
1.7439 & 1.7571 & 0 & 0.513\\
1.7571 & 1.7704 & 1 & 0.478\\
1.7704 & 1.7839 & 1 & 0.475\\
1.7839 & 1.7974 & 0 & 0.440\\
1.7974 & 1.811 & 0 & 0.434\\
1.811 & 1.8246 & 0 & 0.418\\
1.8246 & 1.8384 & 2 & 0.384\\
1.8384 & 1.8523 & 0 & 0.370\\
1.8523 & 1.8662 & 0 & 0.367\\
1.8662 & 1.8803 & 0 & 0.341\\
1.8803 & 1.8944 & 0 & 0.338\\
1.8944 & 1.9086 & 0 & 0.323\\
1.9086 & 1.923 & 0 & 0.294\\
1.923 & 1.9374 & 0 & 0.293\\
1.9374 & 1.9519 & 0 & 0.295\\
1.9519 & 1.9665 & 0 & 0.257\\
1.9665 & 1.9812 & 0 & 0.266\\
1.9812 & 1.996 & 0 & 0.243\\
1.996 & 2.0109 & 0 & 0.241\\
2.0109 & 2.0259 & 0 & 0.226\\
2.0259 & 2.041 & 0 & 0.223\\
2.041 & 2.0561 & 0 & 0.207\\
2.0561 & 2.0714 & 0 & 0.204\\
2.0714 & 2.0868 & 0 & 0.196\\
2.0868 & 2.1023 & 0 & 0.189\\
2.1023 & 2.1179 & 2 & 0.174\\
2.1179 & 2.1336 & 0 & 0.173\\
2.1336 & 2.1494 & 0 & 0.163\\
2.1494 & 2.1653 & 0 & 0.159\\
2.1653 & 2.1813 & 0 & 0.146\\
2.1813 & 2.1974 & 0 & 0.146\\
2.1974 & 2.2136 & 0 & 0.138\\
2.2136 & 2.2299 & 1 & 0.131\\
2.2299 & 2.2463 & 0 & 0.128\\
2.2463 & 2.2629 & 0 & 0.121\\
2.2629 & 2.2795 & 0 & 0.119\\
2.2795 & 2.2963 & 0 & 0.109\\
2.2963 & 2.3131 & 0 & 0.106\\
2.3131 & 2.3301 & 0 & 0.099\\
2.3301 & 2.3472 & 0 & 0.095\\
2.3472 & 2.3643 & 2 & 0.092\\
2.3643 & 2.3816 & 0 & 0.089\\
2.3816 & 2.3991 & 1 & 0.082\\
2.3991 & 2.4166 & 0 & 0.078\\
2.4166 & 2.4342 & 0 & 0.077\\
2.4342 & 2.452 & 0 & 0.071\\
2.452 & 2.4699 & 0 & 0.068\\
2.4699 & 2.4879 & 0 & 0.064\\
2.4879 & 2.506 & 0 & 0.061\\
2.506 & 2.5242 & 0 & 0.059\\
2.5242 & 2.5425 & 0 & 0.054\\
2.5425 & 2.561 & 1 & 0.053\\
2.561 & 2.5796 & 0 & 0.050\\
2.5796 & 2.5983 & 0 & 0.046\\
2.5983 & 2.6171 & 0 & 0.045\\
2.6171 & 2.6361 & 0 & 0.042\\
2.6361 & 2.6551 & 0 & 0.040\\
2.6551 & 2.6743 & 0 & 0.039\\
2.6743 & 2.6937 & 0 & 0.035\\
2.6937 & 2.7131 & 0 & 0.036\\
2.7131 & 2.7327 & 0 & 0.032\\
2.7327 & 2.7524 & 0 & 0.031\\
2.7524 & 2.7722 & 0 & 0.028\\
2.7722 & 2.7922 & 0 & 0.026\\
2.7922 & 2.8123 & 0 & 0.025\\
2.8123 & 2.8325 & 0 & 0.023\\
2.8325 & 2.8529 & 0 & 0.022\\
2.8529 & 2.8734 & 0 & 0.021\\
2.8734 & 2.894 & 0 & 0.019\\
2.894 & 2.9147 & 0 & 0.018\\
2.9147 & 2.9356 & 1 & 0.016\\
2.9356 & 2.9567 & 0 & 0.015\\
2.9567 & 2.9778 & 0 & 0.015\\
2.9778 & 2.9991 & 0 & 0.013\\
2.9991 & 3.0206 & 0 & 0.012\\
3.0206 & 3.0421 & 0 & 0.012\\
3.0421 & 3.0639 & 0 & 0.010\\
3.0639 & 3.0857 & 0 & 0.010\\
3.0857 & 3.1077 & 0 & 0.010\\
3.1077 & 3.1299 & 0 & 0.008\\
3.1299 & 3.1522 & 0 & 0.007\\
3.1522 & 3.1746 & 0 & 0.006\\
3.1746 & 3.1972 & 0 & 0.006\\
3.1972 & 3.2199 & 0 & 0.005\\
3.2199 & 3.2428 & 0 & 0.005\\
3.2428 & 3.2659 & 0 & 0.004\\
3.2659 & 3.289 & 0 & 0.004\\
3.289 & 3.3124 & 0 & 0.003\\
3.3124 & 3.3358 & 0 & 0.003\\
3.3358 & 3.3595 & 0 & 0.002\\
3.3595 & 3.3833 & 0 & 0.002\\
3.3833 & 3.4072 & 0 & 0.001\\
3.4072 & 3.4313 & 0 & 0\\
3.4313 & 3.4556 & 0 & 0\\
3.4556 & 3.48 & 0 & 0\\
3.48 & 3.5046 & 0 & 0\\
3.5046 & 3.5293 & 0 & 0\\
3.5293 & 3.5542 & 0 & 0\\
3.5542 & 3.5792 & 0 & 0\\
3.5792 & 3.6045 & 0 & 0\\
3.6045 & 3.6298 & 0 & 0\\
3.6298 & 3.6554 & 0 & 0\\
3.6554 & 3.6811 & 0 & 0\\
3.6811 & 3.707 & 0 & 0\\
3.707 & 3.733 & 0 & 0\\
3.733 & 3.7593 & 0 & 0\\
3.7593 & 3.7857 & 0 & 0\\
3.7857 & 3.8122 & 0 & 0\\
3.8122 & 3.839 & 0 & 0\\
3.839 & 3.8659 & 0 & 0\\
3.8659 & 3.893 & 0 & 0\\
3.893 & 3.9202 & 0 & 0\\
3.9202 & 3.9477 & 0 & 0\\
3.9477 & 3.9753 & 0 & 0\\
3.9753 & 4.0031 & 0 & 0\\
4.0031 & 4.031 & 0 & 0\\
4.031 & 4.0592 & 0 & 0\\
4.0592 & 4.0875 & 0 & 0\\
4.0875 & 4.1161 & 0 & 0\\
4.1161 & 4.1448 & 0 & 0\\
4.1448 & 4.1737 & 0 & 0\\
4.1737 & 4.2028 & 0 & 0\\
4.2028 & 4.2321 & 0 & 0\\
4.2321 & 4.2615 & 0 & 0\\
4.2615 & 4.2912 & 0 & 0\\
4.2912 & 4.321 & 0 & 0\\
4.321 & 4.3511 & 0 & 0\\
4.3511 & 4.3813 & 0 & 0\\
4.3813 & 4.4118 & 0 & 0\\
4.4118 & 4.4424 & 0 & 0\\
4.4424 & 4.4732 & 0 & 0\\
4.4732 & 4.5043 & 0 & 0\\
4.5043 & 4.5355 & 0 & 0\\
4.5355 & 4.5669 & 0 & 0\\
4.5669 & 4.5986 & 0 & 0\\
4.5986 & 4.6304 & 0 & 0\\
4.6304 & 4.6625 & 0 & 0\\
4.6625 & 4.6948 & 0 & 0\\
4.6948 & 4.7272 & 0 & 0\\
4.7272 & 4.7599 & 0 & 0\\
4.7599 & 4.7928 & 0 & 0\\
4.7928 & 4.8259 & 0 & 0\\
4.8259 & 4.8593 & 0 & 0\\
4.8593 & 4.8928 & 0 & 0\\
4.8928 & 4.9266 & 0 & 0\\
4.9266 & 4.9606 & 0 & 0\\
4.9606 & 4.9948 & 0 & 0\\
4.9948 & 5.0292 & 0 & 0\\
\bottomrule
\bottomrule
\end{longtable}

\begin{longtable}{llll}
\caption{Expected and observed event yields in the dimuon channel, directly corresponding to the non-linear binning presented in Figure~\ref{plots_BH_mm}. The expected yield is given up to at most 4 digit precision.} \label{mm_search_table} \\
\toprule
\toprule
Lower edge [\TeV] & Upper edge [\TeV] & Data [N] & Total Background [N] \\
\midrule
0.12016 & 0.12264 & 18410 & 18200\\
0.12264 & 0.12518 & 16432 & 16330\\
0.12518 & 0.12779 & 14813 & 14840\\
0.12779 & 0.13047 & 13545 & 13540\\
0.13047 & 0.13322 & 12246 & 12410\\
0.13322 & 0.13605 & 11539 & 11430\\
0.13605 & 0.13895 & 10520 & 10540\\
0.13895 & 0.14193 & 9737 & 9778\\
0.14193 & 0.145 & 8911 & 9085\\
0.145 & 0.14816 & 8435 & 8435\\
0.14816 & 0.1514 & 7944 & 7904\\
0.1514 & 0.15474 & 7448 & 7365\\
0.15474 & 0.15817 & 6843 & 6907\\
0.15817 & 0.16171 & 6542 & 6490\\
0.16171 & 0.16535 & 6012 & 6107\\
0.16535 & 0.1691 & 5748 & 5707\\
0.1691 & 0.17296 & 5411 & 5364\\
0.17296 & 0.17695 & 5045 & 5023\\
0.17695 & 0.18106 & 4772 & 4739\\
0.18106 & 0.18529 & 4354 & 4437\\
0.18529 & 0.18966 & 4200 & 4209\\
0.18966 & 0.19417 & 3950 & 3968\\
0.19417 & 0.19883 & 3786 & 3740\\
0.19883 & 0.20364 & 3580 & 3530\\
0.20364 & 0.20862 & 3346 & 3290\\
0.20862 & 0.21376 & 3061 & 3069\\
0.21376 & 0.21908 & 2990 & 2902\\
0.21908 & 0.22458 & 2732 & 2716\\
0.22458 & 0.23027 & 2543 & 2546\\
0.23027 & 0.23617 & 2250 & 2387\\
0.23617 & 0.24229 & 2279 & 2238\\
0.24229 & 0.24862 & 2056 & 2109\\
0.24862 & 0.2552 & 1995 & 1973\\
0.2552 & 0.26202 & 1833 & 1844\\
0.26202 & 0.2691 & 1752 & 1711\\
0.2691 & 0.27646 & 1592 & 1596\\
0.27646 & 0.28411 & 1435 & 1476\\
0.28411 & 0.29207 & 1390 & 1376\\
0.29207 & 0.30036 & 1242 & 1294\\
0.30036 & 0.30899 & 1192 & 1198\\
0.30899 & 0.31798 & 1072 & 1113\\
0.31798 & 0.32736 & 1018 & 1038\\
0.32736 & 0.33716 & 948 & 928.1\\
0.33716 & 0.34739 & 879 & 859.4\\
0.34739 & 0.35809 & 788 & 790.5\\
0.35809 & 0.36928 & 710 & 741.1\\
0.36928 & 0.381 & 670 & 670.8\\
0.381 & 0.39328 & 639 & 625.3\\
0.39328 & 0.40617 & 554 & 554.9\\
0.40617 & 0.41971 & 488 & 500.2\\
0.41971 & 0.43394 & 462 & 461.4\\
0.43394 & 0.44892 & 423 & 415.9\\
0.44892 & 0.4647 & 373 & 380.8\\
0.4647 & 0.48135 & 369 & 338.4\\
0.48135 & 0.49893 & 305 & 298.3\\
0.49893 & 0.51752 & 285 & 267.7\\
0.51752 & 0.53721 & 241 & 241.8\\
0.53721 & 0.55808 & 208 & 215.6\\
0.55808 & 0.58025 & 184 & 185.6\\
0.58025 & 0.60383 & 182 & 166.6\\
0.60383 & 0.62894 & 147 & 147.7\\
0.62894 & 0.65575 & 137 & 130.5\\
0.65575 & 0.68441 & 109 & 113.4\\
0.68441 & 0.71511 & 88 & 96.05\\
0.71511 & 0.74806 & 74 & 83.54\\
0.74806 & 0.78351 & 62 & 70.07\\
0.78351 & 0.82173 & 58 & 59.76\\
0.82173 & 0.86303 & 51 & 51.36\\
0.86303 & 0.90778 & 42 & 42.21\\
0.90778 & 0.95641 & 34 & 37.59\\
0.95641 & 1.0094 & 31 & 31.15\\
1.0094 & 1.0673 & 27 & 26\\
1.0673 & 1.1308 & 20 & 20.31\\
1.1308 & 1.2007 & 16 & 15.91\\
1.2007 & 1.2779 & 13 & 13.11\\
1.2779 & 1.3636 & 13 & 10.92\\
1.3636 & 1.459 & 12 & 6.931\\
1.459 & 1.5659 & 4 & 5.6\\
1.5659 & 1.6861 & 4 & 4.324\\
1.6861 & 1.8222 & 2 & 2.72\\
1.8222 & 1.9772 & 3 & 2.067\\
1.9772 & 2.1548 & 1 & 1.374\\
2.1548 & 2.3599 & 0 & 0.882\\
2.3599 & 2.5987 & 0 & 0.570\\
2.5987 & 2.8794 & 0 & 0.333\\
2.8794 & 3.2131 & 0 & 0.188\\
3.2131 & 3.6149 & 0 & 0.104\\
3.6149 & 4.1068 & 0 & 0.049\\
4.1068 & 4.7222 & 0 & 0.025\\
4.7222 & 5.5164 & 0 & 0.010\\
\bottomrule
\bottomrule
\end{longtable}
}

\section*{Acknowledgements}


We thank CERN for the very successful operation of the LHC, as well as the
support staff from our institutions without whom ATLAS could not be
operated efficiently.

We acknowledge the support of ANPCyT, Argentina; YerPhI, Armenia; ARC, Australia; BMWFW and FWF, Austria; ANAS, Azerbaijan; SSTC, Belarus; CNPq and FAPESP, Brazil; NSERC, NRC and CFI, Canada; CERN; CONICYT, Chile; CAS, MOST and NSFC, China; COLCIENCIAS, Colombia; MSMT CR, MPO CR and VSC CR, Czech Republic; DNRF and DNSRC, Denmark; IN2P3-CNRS, CEA-DSM/IRFU, France; SRNSF, Georgia; BMBF, HGF, and MPG, Germany; GSRT, Greece; RGC, Hong Kong SAR, China; ISF, I-CORE and Benoziyo Center, Israel; INFN, Italy; MEXT and JSPS, Japan; CNRST, Morocco; NWO, Netherlands; RCN, Norway; MNiSW and NCN, Poland; FCT, Portugal; MNE/IFA, Romania; MES of Russia and NRC KI, Russian Federation; JINR; MESTD, Serbia; MSSR, Slovakia; ARRS and MIZ\v{S}, Slovenia; DST/NRF, South Africa; MINECO, Spain; SRC and Wallenberg Foundation, Sweden; SERI, SNSF and Cantons of Bern and Geneva, Switzerland; MOST, Taiwan; TAEK, Turkey; STFC, United Kingdom; DOE and NSF, United States of America. In addition, individual groups and members have received support from BCKDF, the Canada Council, CANARIE, CRC, Compute Canada, FQRNT, and the Ontario Innovation Trust, Canada; EPLANET, ERC, ERDF, FP7, Horizon 2020 and Marie Sklodowska-Curie Actions, European Union; Investissements d'Avenir Labex and Idex, ANR, R{\'e}gion Auvergne and Fondation Partager le Savoir, France; DFG and AvH Foundation, Germany; Herakleitos, Thales and Aristeia programmes co-financed by EU-ESF and the Greek NSRF; BSF, GIF and Minerva, Israel; BRF, Norway; CERCA Programme Generalitat de Catalunya, Generalitat Valenciana, Spain; the Royal Society and Leverhulme Trust, United Kingdom.

The crucial computing support from all WLCG partners is acknowledged gratefully, in particular from CERN, the ATLAS Tier-1 facilities at TRIUMF (Canada), NDGF (Denmark, Norway, Sweden), CC-IN2P3 (France), KIT/GridKA (Germany), INFN-CNAF (Italy), NL-T1 (Netherlands), PIC (Spain), ASGC (Taiwan), RAL (UK) and BNL (USA), the Tier-2 facilities worldwide and large non-WLCG resource providers. Major contributors of computing resources are listed in Ref.~\cite{ATL-GEN-PUB-2016-002}.



\printbibliography

\newpage
\begin{flushleft}
{\Large The ATLAS Collaboration}

\bigskip

M.~Aaboud$^\textrm{\scriptsize 137d}$,
G.~Aad$^\textrm{\scriptsize 88}$,
B.~Abbott$^\textrm{\scriptsize 115}$,
O.~Abdinov$^\textrm{\scriptsize 12}$$^{,*}$,
B.~Abeloos$^\textrm{\scriptsize 119}$,
S.H.~Abidi$^\textrm{\scriptsize 161}$,
O.S.~AbouZeid$^\textrm{\scriptsize 139}$,
N.L.~Abraham$^\textrm{\scriptsize 151}$,
H.~Abramowicz$^\textrm{\scriptsize 155}$,
H.~Abreu$^\textrm{\scriptsize 154}$
% ...
\bigskip
\\
% ...
$^{12}$ Institute of Physics, Azerbaijan Academy of Sciences, Baku, Azerbaijan\\
% ...
$^{88}$ CPPM, Aix-Marseille Universit{\'e} and CNRS/IN2P3, Marseille, France\\
% ...
$^{115}$ Homer L. Dodge Department of Physics and Astronomy, University of Oklahoma, Norman OK, United States of America\\
% ...
$^{119}$ LAL, Univ. Paris-Sud, CNRS/IN2P3, Universit{\'e} Paris-Saclay, Orsay, France\\
% ...
$^{137}$ $^{(a)}$ Facult{\'e} des Sciences Ain Chock, R{\'e}seau Universitaire de Physique des Hautes Energies - Universit{\'e} Hassan II, Casablanca; $^{(b)}$ Centre National de l'Energie des Sciences Techniques Nucleaires, Rabat; $^{(c)}$ Facult{\'e} des Sciences Semlalia, Universit{\'e} Cadi Ayyad, LPHEA-Marrakech; $^{(d)}$ Facult{\'e} des Sciences, Universit{\'e} Mohamed Premier and LPTPM, Oujda; $^{(e)}$ Facult{\'e} des sciences, Universit{\'e} Mohammed V, Rabat, Morocco\\
% ...
$^{139}$ Santa Cruz Institute for Particle Physics, University of California Santa Cruz, Santa Cruz CA, United States of America\\
% ...
$^{151}$ Department of Physics and Astronomy, University of Sussex, Brighton, United Kingdom\\
% ...
$^{154}$ Department of Physics, Technion: Israel Institute of Technology, Haifa, Israel\\
$^{155}$ Raymond and Beverly Sackler School of Physics and Astronomy, Tel Aviv University, Tel Aviv, Israel\\
% ...
$^{161}$ Department of Physics, University of Toronto, Toronto ON, Canada\\
% ...
$^{a}$ Also at Department of Physics, King's College London, London, United Kingdom\\
$^{b}$ Also at Institute of Physics, Azerbaijan Academy of Sciences, Baku, Azerbaijan\\
$^{c}$ Also at Novosibirsk State University, Novosibirsk, Russia\\
$^{d}$ Also at TRIUMF, Vancouver BC, Canada\\
$^{e}$ Also at Department of Physics {\&} Astronomy, University of Louisville, Louisville, KY, United States of America\\
% ...
$^{*}$ Deceased
\end{flushleft}


\end{document}